\documentclass{aa}
\usepackage{graphicx}
\usepackage[varg]{txfonts}
\usepackage{natbib}

\newcommand{\zav}[1]{\left(#1\right)}

\newlength\staretab
\newcommand{\Teff}{\ensuremath{T_\mathrm{eff}}}
\newcommand{\logg}{\ensuremath{\log g}}

\newcommand\msr{\ensuremath{M_\odot\,\text{yr}^{-1}}}
\newcommand\kms{\ensuremath{\text{km}\,\text{s}^{-1}}}

\newcommand\ergs{\ensuremath{\text{erg}\,\text{s}^{-1}}}

\newcommand\hvezdaj{\mbox{CD--46 8926}}
\newcommand\hvezdad{\mbox{CD--51 11879}}

\begin{document}

\title{Hot subdwarf wind models with accurate abundances}
\subtitle{II. Helium-dominated merger products CD--46 8926 and CD--51 11879\thanks{Based
on observations collected at the European Southern Observatory, Paranal, Chile
(ESO programme 097.D-0540(A)).}}

\author{J.~Krti\v{c}ka\inst{1} \and I.~Krti\v{c}kov\'a\inst{1} \and
J.~Jan\'\i k\inst{1}
\and P.~N\'emeth\inst{2,3} \and J.~Kub\'at\inst{2} \and M.~Vu\v ckovi\'c\inst{4}}

\institute{\'Ustav teoretick\'e fyziky a astrofyziky, Masarykova univerzita,
           Kotl\'a\v rsk\' a 2, CZ-611\,37 Brno, Czech
           Republic
           \and
           Astronomick\'y \'ustav, Akademie v\v{e}d \v{C}esk\'e
           republiky, Fri\v{c}ova 298, CZ-251 65 Ond\v{r}ejov, Czech Republic
           \and
           Astroserver.org, F\H{o} t\'{e}r 1, 8533 Malomsok, Hungary
           \and
           Instituto de F\'\i sica y Astronom\'\i a, Facultad de Ciencias,
           Universidad de Valpara\'\i so, Gran Breta\~na 1111, Playa Ancha,
           2360102, Valpara\'\i so, Chile 
}

\date{Received}

\abstract
{Helium-dominated subdwarfs are core helium burning stars stripped of
their envelope. The nuclear evolution of these stars alters surface abundances.
Modified abundances impact the strength of the stellar wind.}
{We aim to
understand the influence of modified surface abundances on the strength of the
stellar wind in the helium-dominated subdwarfs CD--46 8926 and CD--51 11879.
A modified wind strength could resolve the problem with the X-ray emission of these
stars, as the expected X-ray luminosity of both stars is significantly higher than
the upper limit determined from observations.}
{We used our own optical spectroscopy combined
with archival ultraviolet spectroscopy and photometry to derive basic parameters
and surface abundances of selected subdwarfs. The resulting parameterst served as
input for the METUJE stellar wind code, which predicts the wind structure of these stars.
We compared the derived wind parameters with the predictions derived for solar
abundances.}{The optical analysis showed that both subdwarfs have effective
temperatures in excess of $60\,$kK and a strong overabundance of carbon in the
case of CD--46 8926 and nitrogen in the case of CD--51 11879. We interpret the
abundance patterns as being a result of enrichment by the products of nuclear
reactions. The modified abundances reduce the wind mass-loss rate by tens of
percent. The reduction improves the predicted wind line profiles in comparison
to observations. The change in helium abundance does not have a strong effect on
the wind parameters. As a result of a lower estimated bolometric luminosity and
mass-loss rate and a larger distance, the expected X-ray luminosities become lower
and agree with observational upper limits.}
{The nucleosynthesis does not
significantly alter the strength of the wind of hot subdwarfs, but the inclusion of
proper stellar parameters improves the agreement with observational wind
characteristics. Our analysis indicates that subdwarfs overabundant in helium
are typically able to launch wind. This conclusion is supported by data gathered
for thousands of subdwarfs from the literature, which shows that subdwarfs overabundant in helium 
avoid the region in the Kiel diagram where the winds are predicted to
be absent. This can be interpreted in terms of the gravitational settling of helium,
which is suppressed by the winds.}



\maketitle

\section{Introduction}

The stellar wind of hot stars is accelerated by the absorption of radiation in
line transitions of heavy elements such as carbon, silicon, and iron
\citep{lusol,cak}. As a result of this, the amount of mass lost by the wind per
unit of time, that is, the wind mass-loss rate, varies as a function of the
atmospheric metal content. Consequently, for a given set of stellar parameters, the
mass-loss rate ranges from zero for a pristine hydrogen-helium composition of
Population III stars \citep{bezvi,cnovit} to values by a factor of few higher
than the solar-metallicity rates in a metal-rich environment \citep{pulboh}.

The variation of the stellar wind mass-loss rate with metallicity is typically
expressed as a function of only one parameter, which is the mass fraction of
heavy elements \citep{vinca,gmmcak,bjorvyv}. This approach provides reasonable
results not only for the Magellanic Clouds \citep{hezkysedi,marcoz} but also
for many other galaxies of the Local Group. However, simple metallicity scaling
may break down once the chemical composition markedly differs from scaled solar
chemical composition. This may happen particularly in later evolutionary phases,
when products of nuclear burning appear on the stellar surface.

\begin{table*}[t]
\caption{Spectra used for the analysis.}
\centering
\label{pozor}
\begin{tabular}{lcccc}
\hline
Star  & Instrument & Spectrum & Domain [\AA] & JD$-2\,400\,000$ \\
\hline
\hvezdaj
& UVES@UT2 & UVES.2016-07-16T00:37:42.207 & 3752--4980 & 57585.5349\\
& IUE/LWR & 11406 & 1850--3300 & 44840.1166\\
& IUE/SWP & 18055 & 1150--1980 & 45234.5364\\
& FUSE & z901550100000 & 900--1190 & 52452.0298 \\
\hline
\hvezdad
& UVES@UT2 & UVES.2016-07-16T03:13:31.450 & 3752--4980 & 57585.6417\\
& IUE/LWR & 13321 & 1850--3300 & 45116.3044\\
& IUE/SWP & 18056 & 1150--1980 & 45234.6662\\
& FUSE & e045020100000 & 900--1190 & 53263.2796 \\
\hline
\end{tabular}
\end{table*}

Hot subdwarfs belong to stars in such a later evolutionary phase, where the
products of nuclear reactions can appear on their surfaces or where the effect
of diffusion can significantly alter the surface chemical composition
\citep[see][for a review]{heberpreh}. Hot subdwarfs are typically core helium
burning objects \citep{guo} that appear below the main sequence in the
Hertzsprung-Russel diagram. There are several evolutionary channels that typically
involve binaries that can lead to hot subluminous objects. A star can be
stripped on the red giant branch due to interaction with a low-mass companion
and ejection of the common envelope \citep{xiong,kramuvol} or due to a stable
Roche-lobe overflow \citep{han}. Signatures of such previous interactions can be
found in a population of hot subdwarfs \citep{saminesami}. A merger of two helium
white dwarfs (or even a white dwarf and a main-sequence star,
\citealt{splybthp}) can also lead to a hot subdwarf, and there is an observational
indication that such objects are helium dominated \citep{geiradprom}. Finally, a
late helium flash can also produce hot subdwarfs \citep{brflasher,sam4}.
The detection of subdwarfs in binaries that may have interacted in the past, for
instance in classical Be stars \citep{bozifiper,petfycan,kourad,robpod}, also
points to a binary origin of some subdwarfs \citep{bepols,shaolin}.

The chemical composition of hot subdwarfs can vary considerably from star to
star. Apart from the abundance of heavy elements, which can be affected by previous
nucleosynthesis and elemental diffusion \citep{dorlah}, even the fraction of the
two most abundant elements, hydrogen and helium, may significantly differ from
solar values. The range of possible helium abundances is truly enormous, from the
helium abundance being lower than the solar value by a few orders of magnitude
\citep{lathbt} up to a helium-dominated chemical composition
\citep{luogailam,nastez,werash}. Low helium abundance is likely caused by helium
gravitational settling \citep{byrnelevit}, while there are indications that
helium overabundance reflects different evolutionary channels that are perhaps connected
with mergers \citep{geiradprom}. But even merger products of helium white dwarfs
may retain some hydrogen on their surfaces, which leads to the formation of hydrogen-rich subdwarfs
\citep{hallvodik}.

Abundance variations influence the wind mass-loss rate, which has several
observational implications. The mass-loss rate is expected to determine the strength of the
X-ray emission, which originates in wind shocks \citep{felpulpal,owomix}. This
may be relevant for hot subdwarfs, some of which show X-ray emission that is
supposed to originate in wind shocks \citep{bufacek,sandroprehled}. Subdwarfs
show peculiarities as the result of atmospheric diffusion \citep{unbuhb,miriri},
which becomes inhibited in the presence of a stronger wind \citep{vasam}.

Despite non-monotonic abundance variations for individual elements, stellar wind
models of subdwarfs are typically calculated either for solar or scaled solar
chemical composition \citep{vinca,snehurka}. To understand the influence of
individual chemical compositions on the wind strength of hot subdwarfs, we initiated
an observational and theoretical program aimed at predicting mass-loss rates for
individual subdwarfs regarding their actual chemical composition. We started with
hydrogen-rich subdwarfs where we have tested the influence of the wind strength
on the X-ray emission and elemental diffusion \citep{esosubwind}. In this second
paper, we
continue our effort by studying two helium-dominated subdwarfs. Both objects have
parameters for which a line-driven wind is expected \citep{snehurka} but still
do not show any X-ray emission \citep{bufacek}.

\section{Spectroscopy and photometry} 

The spectral analysis of our selected helium-dominated subdwarfs is based on optical
spectroscopy obtained by us and through archival UV spectra. Such
a combination of optical and UV spectroscopy can provide reliable stellar
parameters of hot evolved objects \citep[e.g.,][]{lathbt,svlk}. We obtained the
optical spectra using the high-resolution spectrograph UVES $(R=80\,000)$
mounted at the Nasmyth B focus of VLT-UT2 (Kueyen) within our ESO proposal
097.D-0540(A). Both spectra were taken on 16 July 2016 with a total exposure
time of 1500s in the case of \hvezdaj\ and 1260s in the case of \hvezdad.
The spectra were reduced using standard IRAF\footnote{IRAF is distributed by
NOAO, which is operated by AURA, Inc., under a cooperative agreement with the
National Science Foundation.} routines (bias, flat, and wavelength calibration).
The optical spectra cover the spectral region of $3752 - 4980~\AA$. 

We supplemented the optical data with UV spectra derived using the Far
Ultraviolet Spectroscopic Explorer (FUSE) and the International Ultraviolet
Explorer (IUE, SWP and LWR cameras,\footnote{Short-wavelength prime (SWP) and
long-wavelength redundant (LWR) cameras} high-dispersion data) satellites. We
downloaded the processed UV spectra from the MAST archive.\footnote{Mikulski
Archive for Space Telescopes: http://archive.stsci.edu.} The list of all used
spectroscopical observations is given in Table~\ref{pozor}.

The photometry for the analysis of the spectral energy distribution was
downloaded using the Virtual Observatory SED
Analyzer\footnote{http://svo2.cab.inta-csic.es/theory/vosa/} web tool
\citep[VOSA;][]{vosa}. The photometry was based on $ubvy$ magnitudes of
\citet{ernstuvby} and data from TYCHO survey \citep{hogefabe}, the Carlsberg Meridian
Telescope CCD drift scan survey \citep{vosaapass}, the Sloan Digital Sky Survey
\citep[SDSS;][]{vosasloan}, GAIA DR3 data \citep{gaiadr3}, the DENIS survey
\citep{denis}, the Panoramic Survey Telescope and Rapid Response System 1 survey
\citep[Pan-STARRS1;][]{vosapan}, the Two Micron All Sky Survey
\citep[2MASS][]{2mass}, the Visible and Infrared Survey Telescope for Astronomy
survey \citep[VISTA;][]{vosavista}, and Wide-field Infrared Survey Explorer data
\citep[WISE;][]{vosawise}. We adopted the \citet{card} extinction law to correct
the photometric data for interstellar extinction.

\section{Spectral analysis}

The analysis of available spectra was based on the stationary plane-parallel
model atmosphere code TLUSTY \citep[version 200]{tlusty0}. For stars with weak
winds, which are expected in subdwarfs, even the hydrostatic model atmospheres
provide reliable stellar parameters \citep{bourak,kupa}. The models allow for
departures from the local thermodynamical equilibrium (also abbreviated as NLTE
approach), which is crucial in hot subdwarfs. The atomic data, which are mostly
based on the Opacity and Iron Projects \citep{topt,zel0}, were adopted from
\citet{ostar2003}. We used the SYNSPEC code \citep[version 49]{synspec} to
calculate synthetic spectra. The spectra were corrected for the Doppler effect
by means of radial velocity determined from a cross-correlation with theoretical
spectrum as a template \citep{zvezimi}.

The atmospheric parameters, that is, the effective temperatures $T_\text{eff}$,
surface gravities $\log g$, and elemental abundances $\varepsilon_\text{el}$,
were determined by the $\chi^2$ minimization of the difference between observed
and predicted spectra using the simplex method \citep{kobr}. In this work, we give the
abundances as number density ratios relative to helium, that is,
$\varepsilon_\text{el}={N_\text{el}/N_\text{He}}$. For elements not included
in the minimization, we assumed solar \citep{asp09} density ratio
$m_\text{el}N_\text{el}/\zav{m_\text{H}N_\text{H}+m_\text{He}N_\text{He}}$. To
overcome the problems with atmosphere model convergence, we separated the
determination into two steps. First, we calculated a grid of model atmospheres and a synthetic spectra in
$T_\text{eff}$ and $\log g$ with fixed abundances. We minimized the difference
between the observed spectrum and the predicted spectrum, which was interpolated
from the grid, and we derived $T_\text{eff}$ and $\log g$. We used only optical
spectra in this step. As a second step, we calculated a model atmosphere with parameters determined in the previous
step. Based on this model, we minimized the differences between the observed
spectrum and the predicted spectrum calculated for actual abundances. In this
step we used both optical and UV spectra. These steps were repeated until convergence.

\section{Wind modeling}
\label{kapvetmod}

We used the stellar parameters determined from the spectroscopic analysis to
predict wind structure. For this purpose we applied the global wind code METUJE
\citep{cmfkont}. The code solves the structural equations from the hydrostatic
photosphere to the supersonically expanding radiatively driven stellar wind.
This approach provides so-called global (unified) models
\citep{gableri,grahamz,powrdyn,sundyn}. The models are consistently derived from
the solution of hydrodynamical equations in which the radiative force and the
radiative heating terms are determined using the co-moving frame (CMF) radiative
transfer equation. The thermodynamic state of the wind is given by the solution
of the NLTE equations. The atomic data used for the wind modeling are the same
as described in \citet{btvit}. Most of the data is based on the Opacity and Iron
Project calculations \citep{topt,zel0} and the data obtained from the databases
of NIST \citep{nist}, VALD (Piskunov et al. \citeyear{vald1}, Kupka et al.
\citeyear{vald2}), and the Kurucz
website.\footnote{\url{http://kurucz.harvard.edu}} The code assumes a stationary
and spherically symmetric wind. We assumed a smooth wind and neglected wind
inhomogeneities (clumping).

We used the METUJE code to understand the role of the specific chemical composition
of subdwarfs in the driving of the wind. We mainly focused on the wind parameters
that can be tested against observations, that is, the mass-loss rates $\dot M$
and the terminal velocities $v_\infty$. For this purpose, we calculated three
sets of wind models with the same stellar parameters ($\Teff$, $R$, and $M$) but
with different abundances. One set was calculated for a derived stellar chemical
composition (yielding $\dot M$ and $v_\infty$), a second set was calculated for the same
composition of heavy elements but with helium replaced by a solar mixture of
hydrogen and helium (at the same mass fraction, providing $\dot M^\text{H}$ and
$v_\infty^\text{H}$), and a third set was calculated for a solar \citep{asp09} chemical
composition (denoted as $\dot M^\odot$ and $v_\infty^\odot$). We assumed solar
abundance per baryon for the elements whose abundances were not constrained from
spectroscopy.

\section{\hvezdaj}

The helium-rich subdwarf \hvezdaj\ (LSE 153, CPD-46 6542) is considered to be
single based on an absence of photometric and radial velocity variations
\citep{vrtani,vrtanihebra}. A detailed optical NLTE spectral analysis was
presented by \citet{sam15}. From the determined luminosity, the scaling of
\citet{naze} for O stars predicts the X-ray luminosity of
\mbox{$9.9\times10^{30}\,\ergs$}, which is more than a factor of two higher than
the upper limit from Chandra observations $3.8\times10^{30}\,\ergs$
\citep[corrected for {\em Gaia} distance]{bufacek}. Spectropolarimetry has
not revealed any magnetic field strong enough to be detected \citep{bafors}.

\subsection{Determination of stellar parameters}

\begin{table}[t]
\caption{Wavelengths of the strongest lines (in \AA) used for abundance
determination in \hvezdaj.}
\label{hvezdajel}
\begin{tabular}{ll}
\hline
\ion{H}{i}    & H$\beta$, H$\gamma$, H$\delta$ \\
\ion{C}{iii}  & 4068, 4069, 4070, 4650 \\
\ion{C}{iv}   & 3934, 3935, 4219, 4736\\
\ion{N}{iv}   & 1719\\
\ion{N}{v}    & 4604, 4620 \\
\ion{O}{iv}   & 1339,  1343, 1344\\
\ion{Ne}{iv}   & 4283 \\
\ion{Al}{iii} & 4150\\
\ion{Si}{iv}  & 1122, 1128, 1403, 4116 \\
\ion{P}{v}    & 1118, 1128 \\
\ion{S}{v}    & 1122, 1129, 1134, 1502, 1572 \\
\ion{S}{vi}   & 933, 945, 1118, 4162\\
\ion{Fe}{v}   & 1252 -- 1630 \\
\ion{Fe}{vi}  & 1252 -- 1375 \\
\ion{Ni}{v}   & 1140--1165, 1252 -- 1330\\
\ion{Ni}{vi}   & 1140--1165\\
\hline
\end{tabular}
\end{table}

\begin{figure*}
\includegraphics[width=\textwidth]{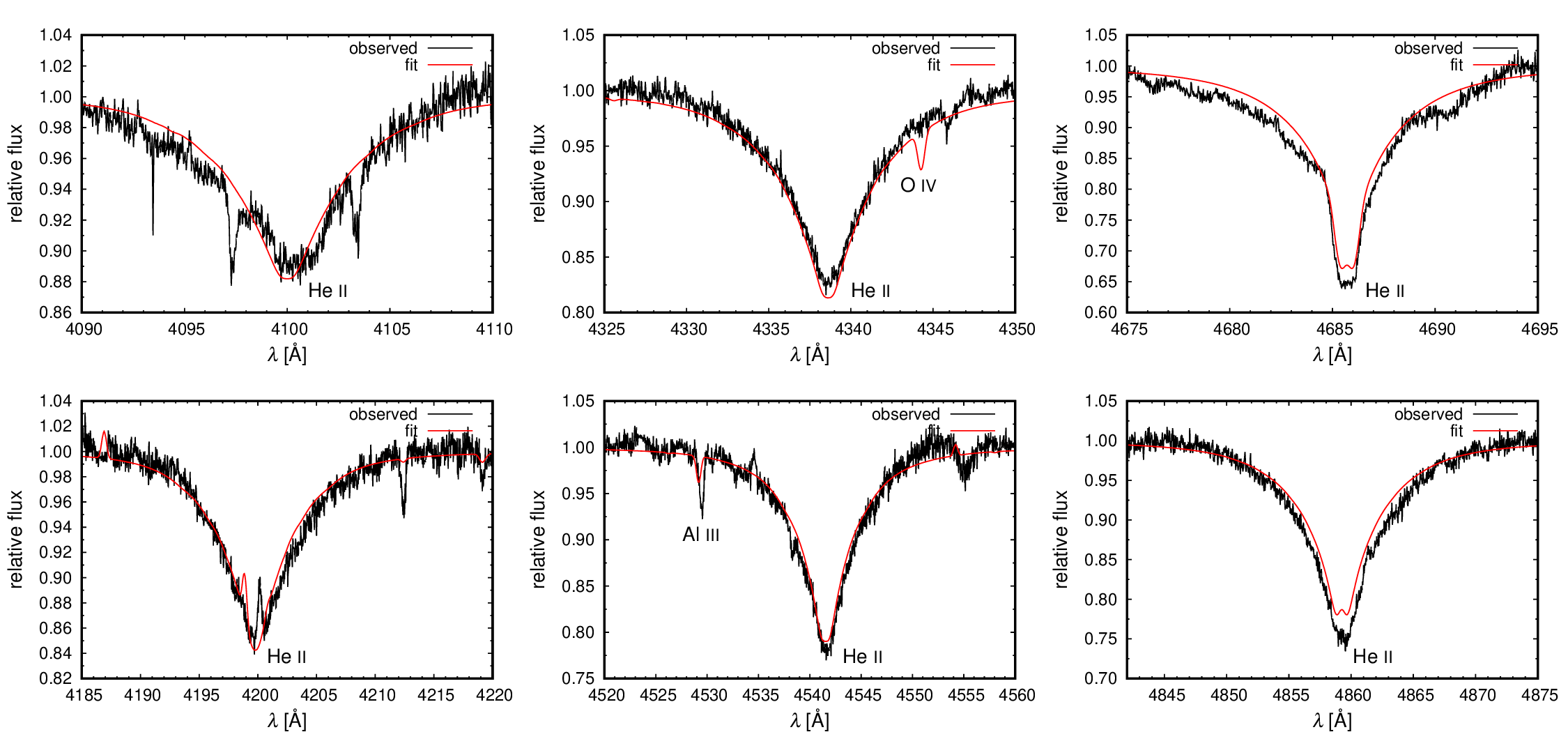}
\caption{Comparison of the best-fit synthetic spectra from SYNSPEC (red line) and UVES
spectra (black line) of \hvezdaj\ in the visual region. We plot the
normalized spectrum as a function of wavelength.}
\label{CD468926uves}
\end{figure*}

\begin{figure*}
\includegraphics[width=\textwidth]{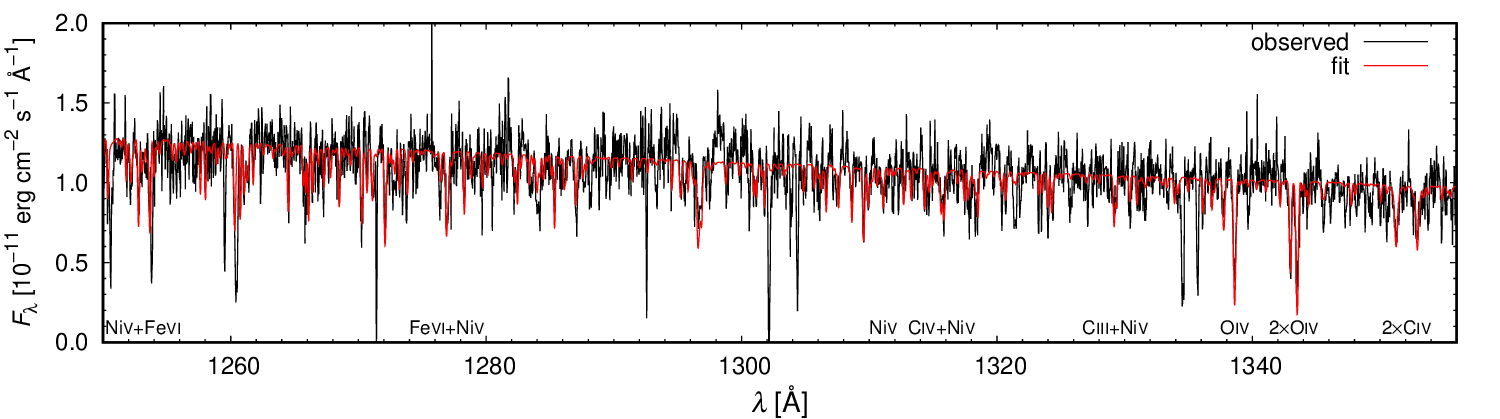}
\caption{Comparison of the best-fit synthetic spectra from SYNSPEC (red line) and IUE
spectra (black line) of \hvezdaj\ in the UV region.}
\label{CD468926uv}
\end{figure*}

The effective temperature and surface gravity were determined from the fit of
optical lines, while we also used FUSE and IUE spectra to derive abundances.
Table~\ref{hvezdajel} lists lines used for abundance determination. Neither the
fit of optical and UV lines was fully satisfactory (see Figs.~\ref{CD468926uves}
and \ref{CD468926uv}), which may possibly reflect missing line opacity in UV. The
optical lines of heavy elements are sensitive to the stellar effective
temperature, and it was not possible to select a single temperature that would fit
all the lines. Most of the derived parameters given in Table~\ref{hvezpar} reasonably agree
with the values determined by \citet{sam15}, with the exception of surface gravity,
which is slightly higher here. The derived abundances roughly correspond to
rescaled solar abundances, with exception of enhanced carbon and nitrogen and
depleted oxygen. Enhanced carbon and nitrogen abundance is typical among hot
subdwarfs \citep{lepsiindex}.

\begin{table}[t]
\caption{Derived parameters of studied stars.}
\label{hvezpar}
\begin{tabular}{lccc}
\hline
Parameter               & \hvezdaj         &  \hvezdad  & Sun \\
\hline
\Teff\ [K]              & $70\,500\pm2000$ & $62\,600\pm3000$ \\
\logg\ [cgs]             & $4.97\pm0.10$    & $5.23\pm0.20$\\
$R$ [$R_\odot$]         & $0.37\pm0.04$    & $0.29\pm0.05$ \\
$M$ [$M_\odot$]         & $0.47\pm0.15$    & $0.52\pm0.31$ \\
$\log(L/L_\odot)$       & $3.48\pm0.11$    & $3.06\pm0.17$ \\
$L_\text{X}$            & $<3.8\times10^{30}$ & $<3.6\times10^{30}$ \\
$\log\varepsilon_\text{H}$  & $<-1.0$      & $<-1.0$& \\
$\log\varepsilon_\text{C}$  & $-2.4\pm0.2$ (0.7)   & $-4.0\pm0.2$ ($-0.9$) & $-3.10$ \\
$\log\varepsilon_\text{N}$  & $-3.3\pm0.5$ (0.4)   & $-3.0\pm0.4$ (0.7) & $-3.70$ \\
$\log\varepsilon_\text{O}$  & $-3.9\pm0.3$ ($-1.1$)& $-4.2\pm0.9$ ($-1.3$) & $-2.84$ \\
$\log\varepsilon_\text{Ne}$ & $-4.7\pm0.3$ ($-1.1$)& $-4.0$ & $-3.60$ \\
$\log\varepsilon_\text{Mg}$ &                      & $-4.2\pm0.3$ ($-0.3$) & $-3.93$ \\
$\log\varepsilon_\text{Al}$ & $-4.3\pm0.3$ ($0.8$) & $-4.9\pm0.3$ (0.1) & $-5.08$ \\
$\log\varepsilon_\text{Si}$ & $-4.6\pm0.5$ ($-0.6$)& $-4.8\pm0.3$ ($-0.8$)& $-4.02$ \\
$\log\varepsilon_\text{P}$  & $-6.6\pm0.3$ ($-0.4$)& $-6.5\pm0.3$ ($-0.4$)& $-6.12$ \\
$\log\varepsilon_\text{S}$  & $-4.3\pm0.3$ (0.1)   & $-4.9\pm0.7$ ($-0.5$) & $-4.41$ \\
$\log\varepsilon_\text{Fe}$ & $-4.7\pm0.4$ ($-0.7$)& $-4.4\pm0.1$ ($-0.4$) & $-4.03$ \\
$\log\varepsilon_\text{Ni}$ & $-5.1\pm0.2$ (0.2)   & $-4.3\pm0.3$ (1.0) & $-5.31$ \\
$v_\text{rad}$              & $-5.0\pm1.0$          & $29.1\pm1.3$ \\
$d$ [pc]                    & $812\pm40$           & $719\pm60$ \\
$\dot M$                    & $2.5\times10^{-9}$   & $4.5\times10^{-10}$ \\
$v_\infty$                  & $1020$               & $1170$ \\
$\dot M^\text{H}$           & $2.4\times10^{-9}$   & $4.3\times10^{-10}$ \\
$v_\infty^\text{H}$         & $930$                & $1180$ \\
$\dot M^\odot$              & $3.4\times10^{-9}$   & $7.0\times10^{-10}$ \\
$v_\infty^\odot$            & $1420$               & $1580$ \\
\hline
\end{tabular}\\
\tablefoot{X-ray luminosities $L_\text{X}$ given in \ergs\ are upper limits
from \citet{bufacek}.
Distances were determined from {\em Gaia} DR3 data
\citep{gaia1,gaia2,gaiadr3}. Solar abundances $\log\varepsilon_\odot$ were taken
from \citet{asp09}. They were rescaled assuming that all hydrogen was converted
into helium. The abundances in parentheses give the value of [X] defined by
\citet{jefkas}. Blank items denote values that were not determined. The radial
velocities were derived from the UVES spectrum. Mass-loss rates (introduced in
Sect.~\ref{kapvetmod}) are given in \msr\ and terminal and radial velocities in
\kms. The abundance of Ne in \hvezdad\ is a mean value determined from
abundances of heavy elements (see Sect.~\ref{kapdvojvit}).}
\end{table}

\begin{figure}
\includegraphics[width=0.5\textwidth]{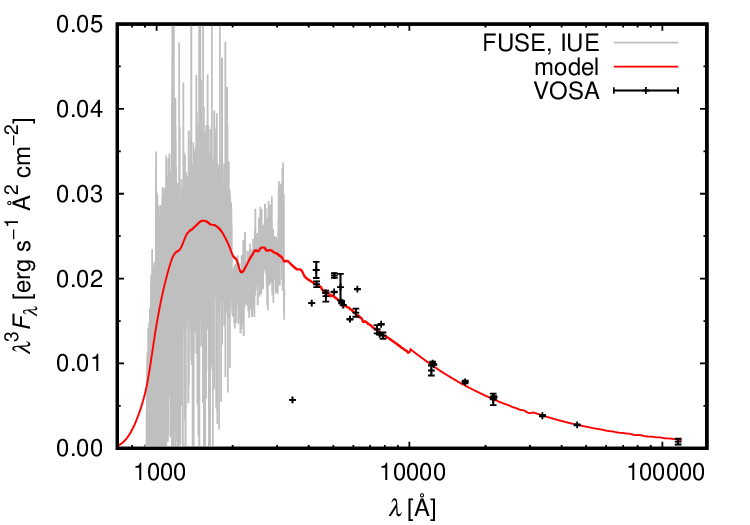}
\caption{Comparison of the predicted spectral energy distribution of \hvezdaj\
from the TLUSTY code
smoothed by a Gaussian filter with a dispersion of $50\,$\AA\ (solid red line) and
observational data derived using VOSA utility \citep{vosa} and from FUSE and IUE
observations.}
\label{cd468926sed}
\end{figure}

We estimated the radius of \hvezdaj\ together with the interstellar reddening
$E(B-V)$ by fitting of the spectral energy distribution
(Fig.~\ref{cd468926sed}). This gave a radius of
$R=0.37\,R_\odot\pm0.04\,R_\odot$ and $E(B-V)=0.07\,$mag, which is slightly
lower than estimated by \citet{peknevrtani}. With spectroscopic surface gravity,
the radius gives the mass as $M=0.47\pm0.15\,M_\odot$.

The radial velocity shift derived from FUSE spectra differs by about
$20\,\kms$ from the shift determined using optical spectroscopy. This is higher
by more than a factor of two than the residual wavelength errors described in the
FUSE Archival Data Handbook.\footnote{https://archive.stsci.edu/fuse/dh.html}
This may indicate the presence of a low-mass companion on the orbit with a period on the order of a year. We also note that an extremely low radial velocity of
$-128.7\pm1.6\,\kms$ determined from the GALAH+ survey \citep{galahp} is most
likely wrong because it is based on the fit of solar abundance spectra.

Helium stars may form via CO+He and He+He white dwarf merger
\citep{hanhehe,jencohe,milbicohe} or by the very late thermal pulse channel.
Carbon overabundance (Table~\ref{hvezpar}) indicates a weak contamination by a
$3\alpha$ reaction following the CNO cycle. The contamination would likely be 
much stronger for CO+He white dwarf merger products, which perhaps points to a
hybrid CO–He+He white dwarf merger origin \citep{jencohe}. Within this scenario,
an sdB subdwarf evolves into a CO–He white dwarf and reignites helium in a shell
after merger with a helium white dwarf. The parameters of \hvezdaj\
(Table~\ref{hvezpar}) correspond to such merger products (Fig.~2 in
\citealt{jencohe}), and the abundances (except Ne) are similar to R~CrB stars
and extreme helium stars \citep{zhaj}. Alternatively, the star could have been
formed from a binary composed of two helium white dwarfs via a composite
merger, which combines a fast merger featuring a hot corona and slow accretion of
the disk \citep{zhaff}. The detected chemical composition may also result from
a delayed helium core flash \citep{brflasher,sam4}.

\subsection{Prediction of wind parameters}

\begin{figure}[t]
\centering
\resizebox{\hsize}{!}{\includegraphics{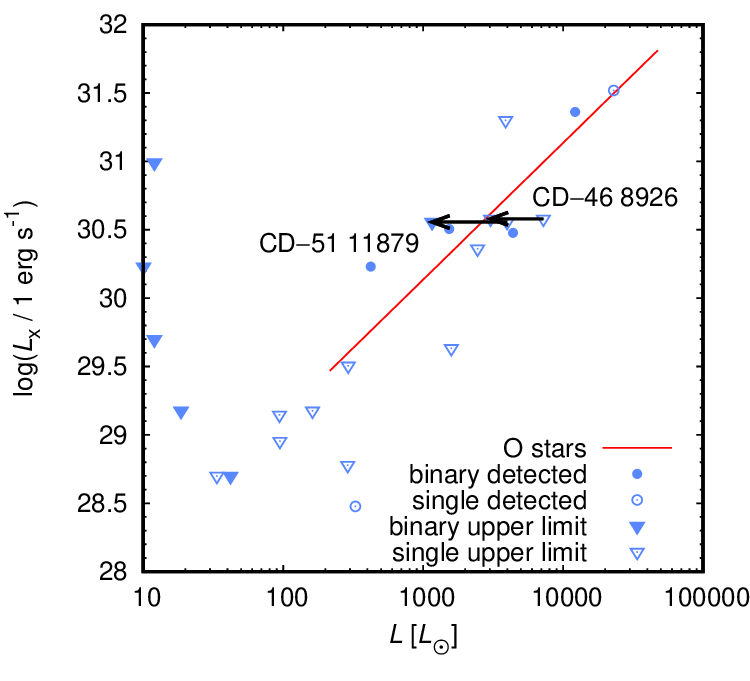}}
\caption{Relation between observed X-ray luminosity and the bolometric
luminosity for subdwarfs (adopted from \citealt{snehurka}). Blue symbols denote
individual subdwarfs with  X-ray detections (circles) and upper limits to X-ray
luminosities (triangles). Filled symbols refer to subdwarfs in binaries, while
empty symbols correspond to single stars. Black arrows with stellar
identifications indicate the shift of the stellar parameters with respect to
previous determinations. The solid red line extrapolates the mean observed relation for
O stars \citep{naze}.}
\label{lxlbolhdbdcd}
\end{figure}

The stellar wind modeling with the determined parameters and abundances predicted a wind
mass-loss rate of $2.5\times10^{-9}\,\msr$ (Table~\ref{hvezpar}). This value is
slightly lower than what was predicted by \citet{snehurka} due to a smaller stellar radius.
As a result of the reduced radius, the X-ray luminosity $
4.1\times10^{30}\,\ergs$ expected for the derived value of bolometric
luminosity (Fig.~\ref{lxlbolhdbdcd}) is close to the observational upper limit of
X-ray luminosity $3.8\times10^{30}\,\ergs$ \citep{bufacek} (Fig.~\ref{lxlbolhdbdcd}). Therefore, with
improved stellar parameters and distance, the models agree much better with X-ray
observations.

The mass-loss rate is not strongly affected by the relative abundance of
hydrogen and helium (Table~\ref{hvezpar}). These elements do not significantly
contribute to the radiative force, and consequently, they influence the mass-loss
rate only indirectly, for example, by a change in the number density of free
electrons or by modification of radiative flux. In our case, a higher helium
abundance leads to a slightly stronger emergent flux at the flux maximum at
about $6\times10^{15}\,\text{s}^{-1}$ and at higher frequencies, where many
resonance lines are located, therefore leading to a slightly higher radiative
force and mass-loss rate. This explains the small difference between $\dot M$ and
$\dot M^\text{H}$ in Table~\ref{hvezpar}. The effect is opposite to that
found by \citet{wrtep} in Wolf-Rayet stars. These stars have large Eddington
factors, and therefore, replacing hydrogen by helium leads to weaker radiative
force due to free electrons and weaker winds.

The difference between the mass-loss rate $\dot M$ predicted for the chemical
composition of \hvezdaj\ and the solar chemical composition $\dot M^\odot$ is caused
by the change in the abundance of individual elements, particularly iron. On
average, the abundance of heavy elements was reduced with respect to the solar
value by $\log\varepsilon-\log\varepsilon_\odot=-0.2$. This abundance reduction
can reproduce the decreased mass-loss rate with the help of Eq.~(1) of
\citet{snehurka}. But the spectral analysis determined the abundances of
only a few of the elements that drive the wind because one-third of the line force
comes from Na and Mg, whose abundances were not estimated within our analysis.
Modified abundances can also explain lower terminal velocities because
some elements that accelerate wind at large speeds, such as O, Ne, and Fe, have
lower abundances.

\begin{figure}[t]
\centering
\resizebox{\hsize}{!}{\includegraphics{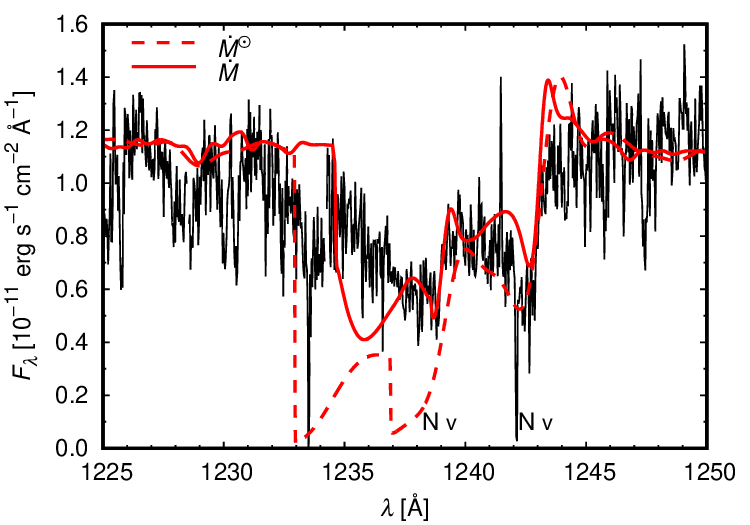}}
\caption{Comparison of observed IUE spectrum of \hvezdaj\ (solid black line)
with emergent flux from wind models METUJE in the region of the \ion{N}{v}
doublet. We plot the emergent fluxes
for mass-loss rate predicted for solar chemical composition (dashed red line)
and mass-loss rate predicted for determined abundances (solid red line).}
\label{cd468926nv}
\end{figure}

When comparing the predicted wind \ion{N}{v} 1239\,\AA\ and 1243\,\AA\ line
profiles with that observed with the IUE satellite (Fig.~\ref{cd468926nv}),
the line profiles predicted for \hvezdaj\ chemical composition are weaker
and agree better with observed spectra than the line profiles predicted for
solar chemical composition. The relative weakness of nitrogen line profiles is
caused by stronger far-UV ionizing radiation coming from helium-dominated
atmospheres that depopulate \ion{N}{v} ion in favor of \ion{N}{vi}.
However, the predicted
\ion{O}{vi} 1032\,\AA\ and 1038\,\AA\ lines show strong P~Cygni line profiles,
while these lines are nearly absent in the FUSE spectra. This could be an indication
of lower oxygen abundance than has been estimated from spectroscopy or the influence of wind
structure on emergent line profiles \citep{chuchcar,clres1}.

\begin{figure*}
\includegraphics[width=\textwidth]{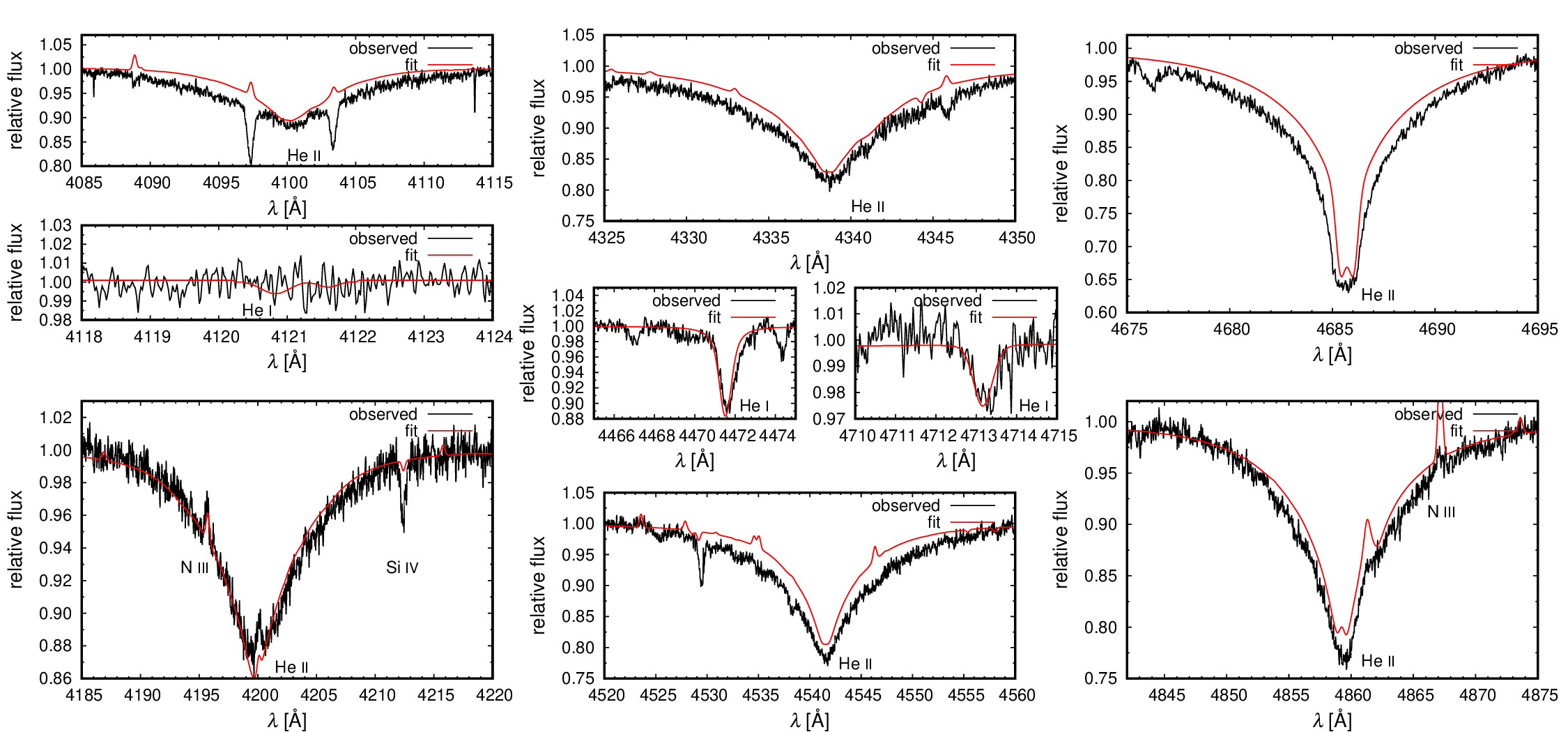}
\caption{Comparison of the best-fit synthetic spectra from SYNSPEC (red line) and UVES
spectra (black line) of \hvezdad\ in the visual region. We plot the
normalized spectrum as a function of wavelength.}
\label{cd5111879uves}
\end{figure*}

\section{\hvezdad}

The hot helium subdwarf \hvezdad\ (LSE 263, ALS 19042) was discovered by
\citet{vrtani}. It was soon recognized that it has a very high effective
temperature \citep{peknevrtani}, which was further supported by optical NLTE
spectral analysis of \citet{sam15}. The star was studied within the
low-resolution optical survey of \citet{nemgalex}. Neither the search for X-ray
emission nor a magnetic field were successful, with a respective upper limit of
$3.6\times10^{30}\,\ergs$ \citep[corrected for {\em Gaia} distance]{bufacek} and
a mean longitudinal magnetic field strength of $\langle
B_z\rangle=360\pm250\,\text{G}$ \citep{bafors}. However, the observational
upper limit of the X-ray luminosity is lower than the estimate of
$5.9\times10^{30}\,\ergs$ based on the scaling of \citet{naze}.

\subsection{Determination of stellar parameters}

\begin{figure*}
\includegraphics[width=\textwidth]{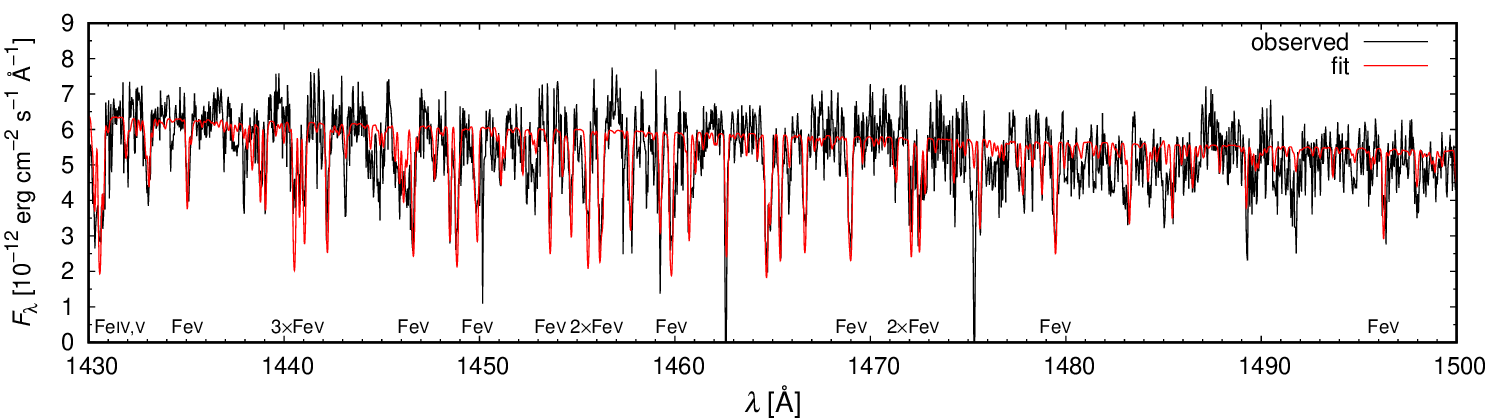}
\caption{Comparison of the best-fit synthetic spectra from SYNSPEC (red line) and IUE
spectra (black line) of \hvezdad\ in the UV region.}
\label{cd5111879uv}
\end{figure*}

\begin{table}[t]
\caption{Wavelengths of the strongest lines (in \AA) used for abundance
determination in \hvezdad.}
\label{cd5111879el}
\begin{tabular}{ll}
\hline
\ion{H}{i}    & H$\beta$, H$\gamma$, H$\delta$ \\
\ion{C}{iii}  & 1175, 1176 \\ 
\ion{C}{iv}   & 1169, 1548, 1551, 4442 \\ 
\ion{N}{iii}  & 3771, 3999, 4004, 4592\\
\ion{N}{iv}   & 1719, 4058, 4512, 4748, 4762\\
\ion{N}{v}    & 4604, 4620 \\
\ion{O}{iii}  &  3755, 3757, 3760 \\
\ion{O}{iv}   & 1339,  1343, 1344\\
\ion{Mg}{ii}  & 4481 \\
\ion{Mg}{iii} & 4463, 4498 \\
\ion{Al}{iii} & 4513\\
\ion{Si}{iv}  & 1122, 1128, 1394, 1403, 4212 \\
\ion{P}{v}    & 1118, 1128 \\
\ion{S}{v}    & 1122, 1129, 1502 \\
\ion{S}{vi}   & 1118 \\
\ion{Fe}{v}   & 930 -- 950, 1030 -- 1080, 1252 -- 1630 \\
\ion{Fe}{vi}  & 1252 -- 1375 \\
\ion{Ni}{v}   & 1110--1180 , 1252 -- 1330\\
\ion{Ni}{vi}  & 1110--1180\\
\hline
\end{tabular}
\end{table}

We determined the effective temperature and surface gravity from the fit of
optical helium lines. Notably, the lines of \ion{He}{i} 4471\,\AA\ and
4713\,\AA\ turned out to be relatively sensitive temperature indicators. To
determine the abundances, we supplemented optical spectra with UV spectra
derived by the FUSE and IUE satellites. The list of lines used for the abundance
analysis of \hvezdad\ is given in Table~\ref{cd5111879el}. The final derived
parameters listed in Table~\ref{hvezpar} agree with the values determined by
\citet{sam15}.

A comparison of observed and fitted spectra in the optical region
(Fig.~\ref{cd5111879uves}) showed that while some lines are nicely fitted, the
other lines slightly disagree. This is perhaps connected with several lines that
appear in the observed UV spectrum but are absent in the predicted spectrum
(Fig.~\ref{cd5111879uv}). Enhanced cooling in the outer parts of the atmosphere
could also lead to the disappearance of some emission lines that appear in the
synthetic spectrum but are missing in the observed spectrum
(Fig.~\ref{cd5111879uves}). Moreover, some lines originate from levels
approximated by superlevels in the model atom. This could lead to an incorrect
estimate of level populations.

\begin{figure}
\includegraphics[width=0.5\textwidth]{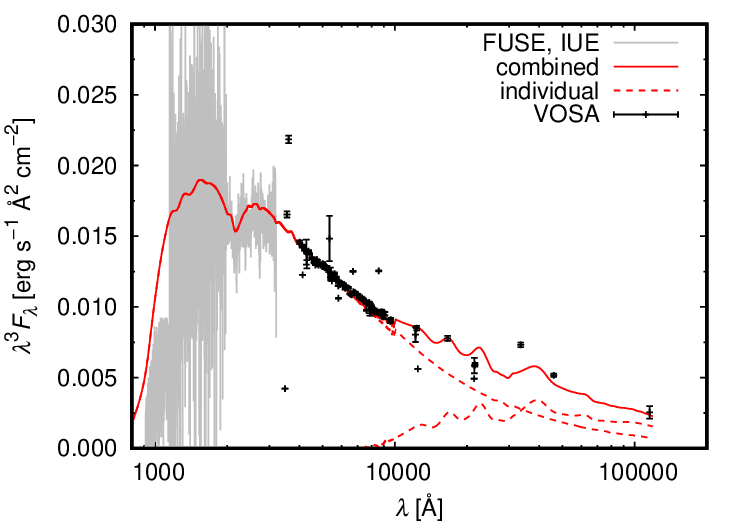}
\caption{Same as Fig.~\ref{cd468926sed} but for \hvezdad\ and featuring
distributions of hot and cool components (dashed lines) and combined
distribution (solid line).}
\label{cd5111879sed}
\end{figure}

{\em Gaia} DR3 data \citep{gaia1,gaia2,gaia3} provide a distance of
$719\pm60\,$pc. Our fit of the observed spectral energy distribution
(Fig.~\ref{cd5111879sed}) gave a stellar radius of
$R=0.29\,R_\odot\pm0.05\,R_\odot$ and an interstellar reddening of $E(B-V)=0.06\,$mag,
which is somewhat lower than estimated by \citet{peknevrtani}. Together with the
spectroscopic surface gravity, the radius gives a mass of $0.52\pm0.31\,M_\odot$.

The RUWE value of 2.4 from the {\em Gaia} data indicates a possible binarity.
Indeed, the spectral energy distribution in Fig.~\ref{cd5111879sed} reveals
the presence of a cooler companion. We combined the subdwarf spectral energy
distribution with solar-metallicity fluxes with $\log g=4$ calculated by
\citet{allmod} and downloaded from the POLLUX database \citep{pollux}. We note
that the spectral energy distribution of the cool companion does not significantly depend on
a particular value of surface gravity. A model with
$T_\text{eff}=2500\,\text{K}$ assuming a radius of $2.2\,R_\odot$ provides a
reasonable fit of the data (Fig.~\ref{cd5111879sed}). This could correspond to a
young low-mass star. Given the absence of radial velocity variations, the
companion might not have affected the evolution of the subdwarf. Moreover,
as a result of its low effective temperature, the companion does not contribute
to the optical spectrum in Fig.~\ref{cd5111879uves}.

\begin{figure}
\includegraphics[width=0.5\textwidth]{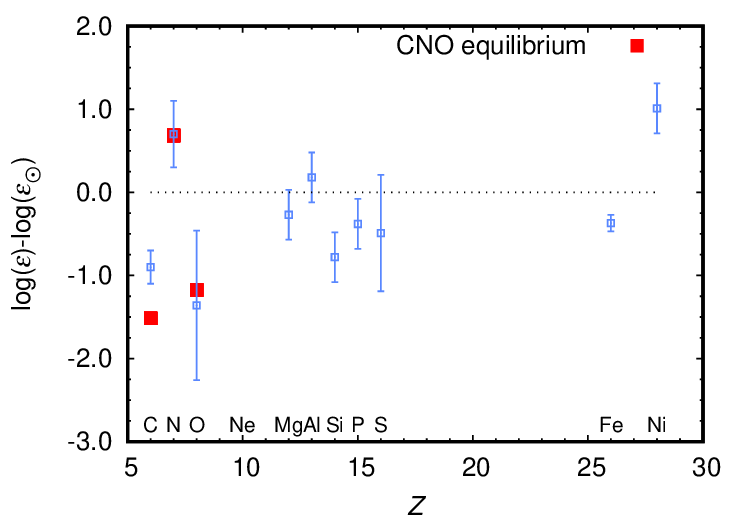}
\caption{Chemical composition of \hvezdad\ relative to the solar abundances
from Table~\ref{hvezpar}. We plot the relative abundances as a function of
atomic number and compare them
to the CNO equilibrium chemical composition.}
\label{cd5111879abun}
\end{figure}

The star shows a helium-dominated chemical composition with an enhanced
abundance of nitrogen and an underabundance of the remaining heavy elements
(Table~\ref{hvezpar}). Figure~\ref{cd5111879abun} shows a comparison of the chemical
composition of \hvezdad\ with the solar chemical composition derived when assuming
that all hydrogen was turned into helium. The \hvezdad\ abundances could
correspond to the CNO processed material, which was suggested to appear on
the surface of a merger product of two helium white dwarfs \citep{saje}. To test
this, we plotted the typical CNO cycle abundances
\citep[Table~25.3]{biblerot} corrected for the mean \hvezdad\ composition
determined using elements from Mg to Fe in Fig.~\ref{cd5111879abun}. The reasonable agreement with
observational values that we found supports the CNO cycle origin of the \hvezdad\ abundances. 

\subsection{Prediction of wind parameters}
\label{kapdvojvit}

The calculation of the wind models revealed that more than half of the
acceleration comes from neon, whose abundance was not determined from
spectroscopy. To alleviate this problem, we adopted a mean determined abundance
(with respect to the solar value) of elements from magnesium to iron as the neon
abundance. This value is given in Table~\ref{hvezpar}.

Table~\ref{hvezpar} shows that the wind parameters predicted using \hvezdad\
abundances and solar abundances do not differ significantly. A large fraction of
the acceleration is connected with optically thick lines that remain optically
thick even at slightly lower abundances, and therefore, the metallicity does
not affect the mass-loss rate significantly. Most of the difference comes from the
lower neon abundance. Also in this case Eq.~(1) of \citet{snehurka} provides
a reasonable estimate of the reduction of the mass-loss rate due to decreased
abundances. Hydrogen and helium do not significantly contribute to the radiative
force. Consequently, the predicted mass-loss rate with the solar mixture of these
elements remain nearly unchanged when hydrogen is replaced by helium
(Table~\ref{hvezpar}). A similar result was found by \citet{vinkolbv} for
luminous blue variables.

The analysis revealed a slightly lower luminosity of the star than determined
previously. As a result, the expected wind X-ray luminosity
($1.6\times10^{30}\,\ergs$, Fig.~\ref{lxlbolhdbdcd}) is below the
observational upper limit $3.6\times10^{30}\,\ergs$ \citep{bufacek}. Therefore, the improved stellar
parameters lead to results that agree with X-ray observations also in the case
of this star.

The inspection of emergent spectra showed that the predicted wind line profiles,
for instance of \ion{O}{vi} 1032\,\AA\ and 1038\,\AA\ lines, are still too
strong when compared with observations. However, the reduced mass-loss
rate improves the agreement between the observed and the predicted spectra in the region
of the 1371\,\AA\ \ion{O}{v} line.

\section{Subdwarfs overabundant in helium in $\Teff$ versus $\log g$ diagrams}

\begin{figure}
\includegraphics[width=0.5\textwidth]{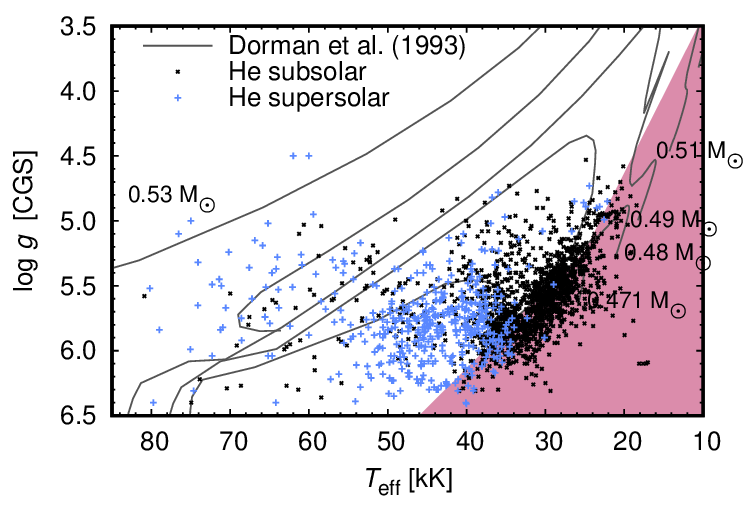}
\caption{Parameters of helium overabundant (blue plus symbols) and helium
underabundant (black crosses) subdwarfs \citep{subkatii} in comparison with the
domain where no wind exists \citep[red area,][]{snehurka}. Overplotted are the
evolutionary tracks of \citet{durman}.}
\label{sitpotvithhe}
\end{figure}

The inclusion of the two additional objects from \citet{esosubwind} showed that
while subdwarfs overabundant in helium have winds, subdwarfs with underabundant
helium do not show winds. This indicates that the existence of the stellar wind
is a necessary condition for appearance of helium overabundance. To further test
this conclusion, we selected subdwarfs with known effective temperatures,
surface gravities, and helium abundances and with estimated uncertainties from the
catalog of \citet{subkatii}. We plotted these parameters in the $\Teff$ versus
$\log g$ (Kiel) diagram (Fig.~\ref{sitpotvithhe}) in comparison with the region
where no wind is predicted \citep{snehurka}. We distinguished subdwarfs that
have a higher abundance of helium than the Sun \citep{asp09} from helium
underabundant subdwarfs.

The comparison in Fig.~\ref{sitpotvithhe} shows that while helium underabundant
subdwarfs appear on both sides of the wind limit, helium overabundant subdwarfs
mostly avoid the region where no wind is predicted. This difference can be
explained in terms of radiative diffusion \citep{miriri,ale3d}. In
hydrodynamically quiet atmospheres without winds, helium sinks due to
gravitational settling. This deprives the surface layers of helium and exposes
hydrogen and radiatively supported heavy elements in the atmosphere. In contrast, strong winds avoid gravitational settling and leave helium in the
atmosphere \citep{vasam,vadog}.

The appearance of a small group of helium overabundant subdwarfs in the region
where the wind is not expected could be connected with the influence of other
parameters, such as mass or metallicity, on the wind limit. Moreover, some stars
could have been deprived of all hydrogen during their previous evolution.

\section{Conclusions}

We used our own optical spectroscopy together with archival UV spectroscopy and
photometry to determine the stellar parameters and surface abundances of two helium
dominated subdwarfs \hvezdaj\ and \hvezdad. Both subdwarfs have an effective
temperature higher than $60\,$kK and an overabundance of either carbon or
nitrogen. There are indications that both stars are binaries.

We computed wind models of both stars from observational stellar parameters.
Both stars are predicted to have relatively weak line-driven wind with mass-loss
rates on the order of $10^{-10}-10^{-9}\,\msr$. The deviation of the chemical
composition from the solar values does not lead to significant changes in the
wind mass-loss rate, not even in the case of helium. These variations in mass-loss
rates induced by modified abundances can be reasonably well
estimated using relations from the literature featuring a mean metallicity.
Modified abundances also lead to lower wind terminal velocities. Wind
models with more realistic abundances predict spectra that better agree with
observations. Moreover, a downward revision of the luminosity shifts the
estimated X-ray luminosity below the observed upper limits.  To a considerable
extent, this was also achieved thanks to a revision of the observational upper
limits based on new {\em Gaia} distances. As a result, the updated stellar
parameters give a predicted X-ray luminosity that is not in conflict with
observations.

Although the chemical compositions of these two stars differ significantly despite
their very similar effective temperatures and surface gravities, it is unlikely
that the surface composition is affected by diffusion (as is typical for cooler
subdwarfs, \citealt{miriri}). Moreover, the wind seen in both stars would likely
wipe out any peculiarities connected with diffusion. Consequently, the
abundances most likely reflect slightly different levels of internal mixing or the
distinct evolutionary paths of each star.

\begin{acknowledgements}
We thank Dr.~Sandro Mereghetti for valuable comments on the manuscript and
Prof.~V.~\v Stefl for the discussion of the cool companion.
Computational resources were provided by the e-INFRA CZ project (ID:90254),
supported by the Ministry of Education, Youth and Sports of the Czech Republic.
PN acknowledges support from the Grant Agency of the Czech Republic
(GA\v{C}R 22-34467S).
The Astronomical
Institute Ond\v{r}ejov is supported by the project RVO:67985815. This
publication makes use of VOSA, developed under the Spanish Virtual Observatory
project supported from the Spanish MINECO through grant AyA2017-84089.
\end{acknowledgements}

\bibliographystyle{aa}
\bibliography{papers}

\begin{thebibliography}{104}
\expandafter\ifx\csname natexlab\endcsname\relax\def\natexlab#1{#1}\fi

\bibitem[{{Alam} {et~al.}(2015){Alam}, {Albareti}, {Allende Prieto}, {Anders},
  {Anderson}, {Anderton}, {Andrews}, {Armengaud}, {Aubourg}, {Bailey}, {Basu},
  {Bautista}, {Beaton}, {Beers}, {Bender}, {Berlind}, {Beutler}, {Bhardwaj},
  {Bird}, {Bizyaev}, {Blake}, {Blanton}, {Blomqvist}, {Bochanski}, {Bolton},
  {Bovy}, {Shelden Bradley}, {Brandt}, {Brauer}, {Brinkmann}, {Brown},
  {Brownstein}, {Burden}, {Burtin}, {Busca}, {Cai}, {Capozzi}, {Carnero
  Rosell}, {Carr}, {Carrera}, {Chambers}, {Chaplin}, {Chen}, {Chiappini},
  {Chojnowski}, {Chuang}, {Clerc}, {Comparat}, {Covey}, {Croft}, {Cuesta},
  {Cunha}, {da Costa}, {Da Rio}, {Davenport}, {Dawson}, {De Lee}, {Delubac},
  {Deshpande}, {Dhital}, {Dutra-Ferreira}, {Dwelly}, {Ealet}, {Ebelke},
  {Edmondson}, {Eisenstein}, {Ellsworth}, {Elsworth}, {Epstein}, {Eracleous},
  {Escoffier}, {Esposito}, {Evans}, {Fan}, {Fern{\'a}ndez-Alvar}, {Feuillet},
  {Filiz Ak}, {Finley}, {Finoguenov}, {Flaherty}, {Fleming}, {Font-Ribera},
  {Foster}, {Frinchaboy}, {Galbraith-Frew}, {Garc{\'\i}a},
  {Garc{\'\i}a-Hern{\'a}ndez}, {Garc{\'\i}a P{\'e}rez}, {Gaulme}, {Ge},
  {G{\'e}nova-Santos}, {Georgakakis}, {Ghezzi}, {Gillespie}, {Girardi},
  {Goddard}, {Gontcho}, {Gonz{\'a}lez Hern{\'a}ndez}, {Grebel}, {Green},
  {Grieb}, {Grieves}, {Gunn}, {Guo}, {Harding}, {Hasselquist}, {Hawley},
  {Hayden}, {Hearty}, {Hekker}, {Ho}, {Hogg}, {Holley-Bockelmann}, {Holtzman},
  {Honscheid}, {Huber}, {Huehnerhoff}, {Ivans}, {Jiang}, {Johnson},
  {Kinemuchi}, {Kirkby}, {Kitaura}, {Klaene}, {Knapp}, {Kneib}, {Koenig},
  {Lam}, {Lan}, {Lang}, {Laurent}, {Le Goff}, {Leauthaud}, {Lee}, {Lee},
  {Licquia}, {Liu}, {Long}, {L{\'o}pez-Corredoira}, {Lorenzo-Oliveira},
  {Lucatello}, {Lundgren}, {Lupton}, {Mack}, {Mahadevan}, {Maia}, {Majewski},
  {Malanushenko}, {Malanushenko}, {Manchado}, {Manera}, {Mao}, {Maraston},
  {Marchwinski}, {Margala}, {Martell}, {Martig}, {Masters}, {Mathur},
  {McBride}, {McGehee}, {McGreer}, {McMahon}, {M{\'e}nard}, {Menzel},
  {Merloni}, {M{\'e}sz{\'a}ros}, {Miller}, {Miralda-Escud{\'e}}, {Miyatake},
  {Montero-Dorta}, {More}, {Morganson}, {Morice-Atkinson}, {Morrison},
  {Mosser}, {Muna}, {Myers}, {Nandra}, {Newman}, {Neyrinck}, {Nguyen},
  {Nichol}, {Nidever}, {Noterdaeme}, {Nuza}, {O'Connell}, {O'Connell},
  {O'Connell}, {Ogando}, {Olmstead}, {Oravetz}, {Oravetz}, {Osumi}, {Owen},
  {Padgett}, {Padmanabhan}, {Paegert}, {Palanque-Delabrouille}, {Pan},
  {Parejko}, {P{\^a}ris}, {Park}, {Pattarakijwanich}, {Pellejero-Ibanez},
  {Pepper}, {Percival}, {P{\'e}rez-Fournon}, {P{\'e}rez-R{\`a}fols},
  {Petitjean}, {Pieri}, {Pinsonneault}, {Porto de Mello}, {Prada}, {Prakash},
  {Price-Whelan}, {Protopapas}, {Raddick}, {Rahman}, {Reid}, {Rich}, {Rix},
  {Robin}, {Rockosi}, {Rodrigues}, {Rodr{\'\i}guez-Torres}, {Roe}, {Ross},
  {Ross}, {Rossi}, {Ruan}, {Rubi{\~n}o-Mart{\'\i}n}, {Rykoff},
  {Salazar-Albornoz}, {Salvato}, {Samushia}, {S{\'a}nchez}, {Santiago},
  {Sayres}, {Schiavon}, {Schlegel}, {Schmidt}, {Schneider}, {Schultheis},
  {Schwope}, {Sc{\'o}ccola}, {Scott}, {Sellgren}, {Seo}, {Serenelli}, {Shane},
  {Shen}, {Shetrone}, {Shu}, {Silva Aguirre}, {Sivarani}, {Skrutskie},
  {Slosar}, {Smith}, {Sobreira}, {Souto}, {Stassun}, {Steinmetz}, {Stello},
  {Strauss}, {Streblyanska}, {Suzuki}, {Swanson}, {Tan}, {Tayar}, {Terrien},
  {Thakar}, {Thomas}, {Thomas}, {Thompson}, {Tinker}, {Tojeiro}, {Troup},
  {Vargas-Maga{\~n}a}, {Vazquez}, {Verde}, {Viel}, {Vogt}, {Wake}, {Wang},
  {Weaver}, {Weinberg}, {Weiner}, {White}, {Wilson}, {Wisniewski},
  {Wood-Vasey}, {Ye`che}, {York}, {Zakamska}, {Zamora}, {Zasowski}, {Zehavi},
  {Zhao}, {Zheng}, {Zhou}, {Zhou}, {Zou}, \& {Zhu}}]{vosasloan}
{Alam}, S., {Albareti}, F.~D., {Allende Prieto}, C., {et~al.} 2015, \apjs, 219,
  12

\bibitem[{{Alecian} \& {Stift}(2021)}]{ale3d}
{Alecian}, G. \& {Stift}, M.~J. 2021, \mnras, 504, 1370

\bibitem[{{Allard} {et~al.}(2012){Allard}, {Homeier}, \& {Freytag}}]{allmod}
{Allard}, F., {Homeier}, D., \& {Freytag}, B. 2012, Philosophical Transactions
  of the Royal Society of London Series A, 370, 2765

\bibitem[{{Asplund} {et~al.}(2009){Asplund}, {Grevesse}, {Sauval}, \&
  {Scott}}]{asp09}
{Asplund}, M., {Grevesse}, N., {Sauval}, A.~J., \& {Scott}, P. 2009, \araa, 47,
  481

\bibitem[{{Bagnulo} {et~al.}(2015){Bagnulo}, {Fossati}, {Landstreet}, \&
  {Izzo}}]{bafors}
{Bagnulo}, S., {Fossati}, L., {Landstreet}, J.~D., \& {Izzo}, C. 2015, \aap,
  583, A115

\bibitem[{{Bayo} {et~al.}(2008){Bayo}, {Rodrigo}, {Barrado Y Navascu{\'e}s},
  {Solano}, {Guti{\'e}rrez}, {Morales-Calder{\'o}n}, \& {Allard}}]{vosa}
{Bayo}, A., {Rodrigo}, C., {Barrado Y Navascu{\'e}s}, D., {et~al.} 2008, \aap,
  492, 277

\bibitem[{{Bj{\"o}rklund} {et~al.}(2023){Bj{\"o}rklund}, {Sundqvist}, {Singh},
  {Puls}, \& {Najarro}}]{bjorvyv}
{Bj{\"o}rklund}, R., {Sundqvist}, J.~O., {Singh}, S.~M., {Puls}, J., \&
  {Najarro}, F. 2023, \aap, 676, A109

\bibitem[{{Bouret} {et~al.}(2003){Bouret}, {Lanz}, {Hillier}, {Heap}, {Hubeny},
  {Lennon}, {Smith}, \& {Evans}}]{bourak}
{Bouret}, J.-C., {Lanz}, T., {Hillier}, D.~J., {et~al.} 2003, \apj, 595, 1182

\bibitem[{{Bo\v zi\'c} {et~al.}(1995){Bo\v zi\'c}, {Harmanec}, {Horn},
  {Koubsky}, {Scholz}, {McDavid}, {Hubert}, \& {Hubert}}]{bozifiper}
{Bo\v zi\'c}, H., {Harmanec}, P., {Horn}, J., {et~al.} 1995, \aap, 304, 235

\bibitem[{{Brands} {et~al.}(2022){Brands}, {de Koter}, {Bestenlehner},
  {Crowther}, {Sundqvist}, {Puls}, {Caballero-Nieves}, {Abdul-Masih},
  {Driessen}, {Garc{\'\i}a}, {Geen}, {Gr{\"a}fener}, {Hawcroft}, {Kaper},
  {Keszthelyi}, {Langer}, {Sana}, {Schneider}, {Shenar}, \& {Vink}}]{hezkysedi}
{Brands}, S.~A., {de Koter}, A., {Bestenlehner}, J.~M., {et~al.} 2022, \aap,
  663, A36

\bibitem[{{Brown} {et~al.}(2001){Brown}, {Sweigart}, {Lanz}, {Landsman}, \&
  {Hubeny}}]{brflasher}
{Brown}, T.~M., {Sweigart}, A.~V., {Lanz}, T., {Landsman}, W.~B., \& {Hubeny},
  I. 2001, \apj, 562, 368

\bibitem[{{Buder} {et~al.}(2021){Buder}, {Sharma}, {Kos}, {Amarsi},
  {Nordlander}, {Lind}, {Martell}, {Asplund}, {Bland-Hawthorn}, {Casey}, {de
  Silva}, {D'Orazi}, {Freeman}, {Hayden}, {Lewis}, {Lin}, {Schlesinger},
  {Simpson}, {Stello}, {Zucker}, {Zwitter}, {Beeson}, {Buck}, {Casagrande},
  {Clark}, {{\v{C}}otar}, {da Costa}, {de Grijs}, {Feuillet}, {Horner},
  {Kafle}, {Khanna}, {Kobayashi}, {Liu}, {Montet}, {Nandakumar}, {Nataf},
  {Ness}, {Spina}, {Tepper-Garc{\'\i}a}, {Ting}, {Traven},
  {Vogrin{\v{c}}i{\v{c}}}, {Wittenmyer}, {Wyse}, {{\v{Z}}erjal}, \& {Galah
  Collaboration}}]{galahp}
{Buder}, S., {Sharma}, S., {Kos}, J., {et~al.} 2021, \mnras, 506, 150

\bibitem[{{Byrne} \& {Jeffery}(2018)}]{byrnelevit}
{Byrne}, C.~M. \& {Jeffery}, C.~S. 2018, \mnras, 481, 3810

\bibitem[{{Cardelli} {et~al.}(1989){Cardelli}, {Clayton}, \& {Mathis}}]{card}
{Cardelli}, J.~A., {Clayton}, G.~C., \& {Mathis}, J.~S. 1989, \apj, 345, 245

\bibitem[{{Castor} {et~al.}(1975){Castor}, {Abbott}, \& {Klein}}]{cak}
{Castor}, J.~I., {Abbott}, D.~C., \& {Klein}, R.~I. 1975, \apj, 195, 157

\bibitem[{{Chambers} {et~al.}(2016){Chambers}, {Magnier}, {Metcalfe},
  {Flewelling}, {Huber}, {Waters}, {Denneau}, {Draper}, {Farrow}, {Finkbeiner},
  {Holmberg}, {Koppenhoefer}, {Price}, {Rest}, {Saglia}, {Schlafly}, {Smartt},
  {Sweeney}, {Wainscoat}, {Burgett}, {Chastel}, {Grav}, {Heasley}, {Hodapp},
  {Jedicke}, {Kaiser}, {Kudritzki}, {Luppino}, {Lupton}, {Monet}, {Morgan},
  {Onaka}, {Shiao}, {Stubbs}, {Tonry}, {White}, {Ba{\~n}ados}, {Bell},
  {Bender}, {Bernard}, {Boegner}, {Boffi}, {Botticella}, {Calamida},
  {Casertano}, {Chen}, {Chen}, {Cole}, {Deacon}, {Frenk}, {Fitzsimmons},
  {Gezari}, {Gibbs}, {Goessl}, {Goggia}, {Gourgue}, {Goldman}, {Grant},
  {Grebel}, {Hambly}, {Hasinger}, {Heavens}, {Heckman}, {Henderson}, {Henning},
  {Holman}, {Hopp}, {Ip}, {Isani}, {Jackson}, {Keyes}, {Koekemoer}, {Kotak},
  {Le}, {Liska}, {Long}, {Lucey}, {Liu}, {Martin}, {Masci}, {McLean}, {Mindel},
  {Misra}, {Morganson}, {Murphy}, {Obaika}, {Narayan}, {Nieto-Santisteban},
  {Norberg}, {Peacock}, {Pier}, {Postman}, {Primak}, {Rae}, {Rai}, {Riess},
  {Riffeser}, {Rix}, {R{\"o}ser}, {Russel}, {Rutz}, {Schilbach}, {Schultz},
  {Scolnic}, {Strolger}, {Szalay}, {Seitz}, {Small}, {Smith}, {Soderblom},
  {Taylor}, {Thomson}, {Taylor}, {Thakar}, {Thiel}, {Thilker}, {Unger},
  {Urata}, {Valenti}, {Wagner}, {Walder}, {Walter}, {Watters}, {Werner},
  {Wood-Vasey}, \& {Wyse}}]{vosapan}
{Chambers}, K.~C., {Magnier}, E.~A., {Metcalfe}, N., {et~al.} 2016, arXiv
  e-prints, arXiv:1612.05560

\bibitem[{{Cross} {et~al.}(2012){Cross}, {Collins}, {Mann}, {Read}, {Sutorius},
  {Blake}, {Holliman}, {Hambly}, {Emerson}, {Lawrence}, \&
  {Noddle}}]{vosavista}
{Cross}, N.~J.~G., {Collins}, R.~S., {Mann}, R.~G., {et~al.} 2012, \aap, 548,
  A119

\bibitem[{{Dorman} {et~al.}(1993){Dorman}, {Rood}, \& {O'Connell}}]{durman}
{Dorman}, B., {Rood}, R.~T., \& {O'Connell}, R.~W. 1993, \apj, 419, 596

\bibitem[{{Dorsch} {et~al.}(2019){Dorsch}, {Latour}, \& {Heber}}]{dorlah}
{Dorsch}, M., {Latour}, M., \& {Heber}, U. 2019, \aap, 630, A130

\bibitem[{{Drilling}(1983)}]{vrtani}
{Drilling}, J.~S. 1983, \apjl, 270, L13

\bibitem[{{Drilling} \& {Heber}(1986)}]{vrtanihebra}
{Drilling}, J.~S. \& {Heber}, U. 1986, in IAU Colloq. 87: Hydrogen Deficient
  Stars and Related Objects, ed. K.~{Hunger}, D.~{Schoenberner}, \&
  N.~{Kameswara Rao}, 23

\bibitem[{{Epchtein} {et~al.}(1999){Epchtein}, {Deul}, {Derriere},
  {Borsenberger}, {Egret}, {Simon}, {Alard}, {Bal{\'a}zs}, {de Batz}, {Cioni},
  {Copet}, {Dennefeld}, {Forveille}, {Fouqu{\'e}}, {Garz{\'o}n}, {Habing},
  {Holl}, {Hron}, {Kimeswenger}, {Lacombe}, {Le Bertre}, {Loup}, {Mamon},
  {Omont}, {Paturel}, {Persi}, {Robin}, {Rouan}, {Tiph{\`e}ne}, {Vauglin}, \&
  {Wagner}}]{denis}
{Epchtein}, N., {Deul}, E., {Derriere}, S., {et~al.} 1999, \aap, 349, 236

\bibitem[{{Evans} {et~al.}(2002){Evans}, {Irwin}, \& {Helmer}}]{vosaapass}
{Evans}, D.~W., {Irwin}, M.~J., \& {Helmer}, L. 2002, \aap, 395, 347

\bibitem[{{Feldmeier} {et~al.}(1997){Feldmeier}, {Puls}, \&
  {Pauldrach}}]{felpulpal}
{Feldmeier}, A., {Puls}, J., \& {Pauldrach}, A.~W.~A. 1997, \aap, 322, 878

\bibitem[{{Gabler} {et~al.}(1989){Gabler}, {Gabler}, {Kudritzki}, {Puls}, \&
  {Pauldrach}}]{gableri}
{Gabler}, R., {Gabler}, A., {Kudritzki}, R.~P., {Puls}, J., \& {Pauldrach}, A.
  1989, \aap, 226, 162

\bibitem[{{Gaia Collaboration} {et~al.}(2018){Gaia Collaboration}, {Brown},
  {Vallenari}, {Prusti}, {de Bruijne}, {Babusiaux}, {Bailer-Jones}, {Biermann},
  {Evans}, {Eyer}, \& et~al.}]{gaia2}
{Gaia Collaboration}, {Brown}, A.~G.~A., {Vallenari}, A., {et~al.} 2018, \aap,
  616, A1

\bibitem[{{Gaia Collaboration} {et~al.}(2021){Gaia Collaboration}, {Brown},
  {Vallenari}, {Prusti}, {de Bruijne}, {Babusiaux}, {Biermann}, {Creevey},
  {Evans}, {Eyer}, {Hutton}, {Jansen}, {Jordi}, {Klioner}, {Lammers},
  {Lindegren}, {Luri}, {Mignard}, {Panem}, {Pourbaix}, {Randich}, {Sartoretti},
  {Soubiran}, {Walton}, {Arenou}, {Bailer-Jones}, {Bastian}, {Cropper},
  {Drimmel}, {Katz}, {Lattanzi}, {van Leeuwen}, {Bakker}, {Cacciari},
  {Casta{\~n}eda}, {De Angeli}, {Ducourant}, {Fabricius}, {Fouesneau},
  {Fr{\'e}mat}, {Guerra}, {Guerrier}, {Guiraud}, {Jean-Antoine Piccolo},
  {Masana}, {Messineo}, {Mowlavi}, {Nicolas}, {Nienartowicz}, {Pailler},
  {Panuzzo}, {Riclet}, {Roux}, {Seabroke}, {Sordo}, {Tanga}, {Th{\'e}venin},
  {Gracia-Abril}, {Portell}, {Teyssier}, {Altmann}, {Andrae}, {Bellas-Velidis},
  {Benson}, {Berthier}, {Blomme}, {Brugaletta}, {Burgess}, {Busso}, {Carry},
  {Cellino}, {Cheek}, {Clementini}, {Damerdji}, {Davidson}, {Delchambre},
  {Dell'Oro}, {Fern{\'a}ndez-Hern{\'a}ndez}, {Galluccio}, {Garc{\'\i}a-Lario},
  {Garcia-Reinaldos}, {Gonz{\'a}lez-N{\'u}{\~n}ez}, {Gosset}, {Haigron},
  {Halbwachs}, {Hambly}, {Harrison}, {Hatzidimitriou}, {Heiter},
  {Hern{\'a}ndez}, {Hestroffer}, {Hodgkin}, {Holl}, {Jan{\ss}en}, {Jevardat de
  Fombelle}, {Jordan}, {Krone-Martins}, {Lanzafame}, {L{\"o}ffler}, {Lorca},
  {Manteiga}, {Marchal}, {Marrese}, {Moitinho}, {Mora}, {Muinonen}, {Osborne},
  {Pancino}, {Pauwels}, {Petit}, {Recio-Blanco}, {Richards}, {Riello},
  {Rimoldini}, {Robin}, {Roegiers}, {Rybizki}, {Sarro}, {Siopis}, {Smith},
  {Sozzetti}, {Ulla}, {Utrilla}, {van Leeuwen}, {van Reeven}, {Abbas}, {Abreu
  Aramburu}, {Accart}, {Aerts}, {Aguado}, {Ajaj}, {Altavilla}, {{\'A}lvarez},
  {{\'A}lvarez Cid-Fuentes}, {Alves}, {Anderson}, {Anglada Varela}, {Antoja},
  {Audard}, {Baines}, {Baker}, {Balaguer-N{\'u}{\~n}ez}, {Balbinot}, {Balog},
  {Barache}, {Barbato}, {Barros}, {Barstow}, {Bartolom{\'e}}, {Bassilana},
  {Bauchet}, {Baudesson-Stella}, {Becciani}, {Bellazzini}, {Bernet}, {Bertone},
  {Bianchi}, {Blanco-Cuaresma}, {Boch}, {Bombrun}, {Bossini}, {Bouquillon},
  {Bragaglia}, {Bramante}, {Breedt}, {Bressan}, {Brouillet}, {Bucciarelli},
  {Burlacu}, {Busonero}, {Butkevich}, {Buzzi}, {Caffau}, {Cancelliere},
  {C{\'a}novas}, {Cantat-Gaudin}, {Carballo}, {Carlucci}, {Carnerero},
  {Carrasco}, {Casamiquela}, {Castellani}, {Castro-Ginard}, {Castro Sampol},
  {Chaoul}, {Charlot}, {Chemin}, {Chiavassa}, {Cioni}, {Comoretto}, {Cooper},
  {Cornez}, {Cowell}, {Crifo}, {Crosta}, {Crowley}, {Dafonte}, {Dapergolas},
  {David}, {David}, {de Laverny}, {De Luise}, {De March}, {De Ridder}, {de
  Souza}, {de Teodoro}, {de Torres}, {del Peloso}, {del Pozo}, {Delbo},
  {Delgado}, {Delgado}, {Delisle}, {Di Matteo}, {Diakite}, {Diener},
  {Distefano}, {Dolding}, {Eappachen}, {Edvardsson}, {Enke}, {Esquej}, {Fabre},
  {Fabrizio}, {Faigler}, {Fedorets}, {Fernique}, {Fienga}, {Figueras},
  {Fouron}, {Fragkoudi}, {Fraile}, {Franke}, {Gai}, {Garabato},
  {Garcia-Gutierrez}, {Garc{\'\i}a-Torres}, {Garofalo}, {Gavras}, {Gerlach},
  {Geyer}, {Giacobbe}, {Gilmore}, {Girona}, {Giuffrida}, {Gomel}, {Gomez},
  {Gonzalez-Santamaria}, {Gonz{\'a}lez-Vidal}, {Granvik},
  {Guti{\'e}rrez-S{\'a}nchez}, {Guy}, {Hauser}, {Haywood}, {Helmi}, {Hidalgo},
  {Hilger}, {H{\l}adczuk}, {Hobbs}, {Holland}, {Huckle}, {Jasniewicz},
  {Jonker}, {Juaristi Campillo}, {Julbe}, {Karbevska}, {Kervella}, {Khanna},
  {Kochoska}, {Kontizas}, {Kordopatis}, {Korn}, {Kostrzewa-Rutkowska},
  {Kruszy{\'n}ska}, {Lambert}, {Lanza}, {Lasne}, {Le Campion}, {Le Fustec},
  {Lebreton}, {Lebzelter}, {Leccia}, {Leclerc}, {Lecoeur-Taibi}, {Liao},
  {Licata}, {Lindstr{\o}m}, {Lister}, {Livanou}, {Lobel}, {Madrero Pardo},
  {Managau}, {Mann}, {Marchant}, {Marconi}, {Marcos Santos}, {Marinoni},
  {Marocco}, {Marshall}, {Martin Polo}, {Mart{\'\i}n-Fleitas}, {Masip},
  {Massari}, {Mastrobuono-Battisti}, {Mazeh}, {McMillan}, {Messina},
  {Michalik}, {Millar}, {Mints}, {Molina}, {Molinaro}, {Moln{\'a}r},
  {Montegriffo}, {Mor}, {Morbidelli}, {Morel}, {Morris}, {Mulone}, {Munoz},
  {Muraveva}, {Murphy}, {Musella}, {Noval}, {Ord{\'e}novic}, {Orr{\`u}},
  {Osinde}, {Pagani}, {Pagano}, {Palaversa}, {Palicio}, {Panahi}, {Pawlak},
  {Pe{\~n}alosa Esteller}, {Penttil{\"a}}, {Piersimoni}, {Pineau}, {Plachy},
  {Plum}, {Poggio}, {Poretti}, {Poujoulet}, {Pr{\v{s}}a}, {Pulone}, {Racero},
  {Ragaini}, {Rainer}, {Raiteri}, {Rambaux}, {Ramos}, {Ramos-Lerate}, {Re
  Fiorentin}, {Regibo}, {Reyl{\'e}}, {Ripepi}, {Riva}, {Rixon}, {Robichon},
  {Robin}, {Roelens}, {Rohrbasser}, {Romero-G{\'o}mez}, {Rowell}, {Royer},
  {Rybicki}, {Sadowski}, {Sagrist{\`a} Sell{\'e}s}, {Sahlmann}, {Salgado},
  {Salguero}, {Samaras}, {Sanchez Gimenez}, {Sanna}, {Santove{\~n}a},
  {Sarasso}, {Schultheis}, {Sciacca}, {Segol}, {Segovia}, {S{\'e}gransan},
  {Semeux}, {Shahaf}, {Siddiqui}, {Siebert}, {Siltala}, {Slezak}, {Smart},
  {Solano}, {Solitro}, {Souami}, {Souchay}, {Spagna}, {Spoto}, {Steele},
  {Steidelm{\"u}ller}, {Stephenson}, {S{\"u}veges}, {Szabados}, {Szegedi-Elek},
  {Taris}, {Tauran}, {Taylor}, {Teixeira}, {Thuillot}, {Tonello}, {Torra},
  {Torra}, {Turon}, {Unger}, {Vaillant}, {van Dillen}, {Vanel}, {Vecchiato},
  {Viala}, {Vicente}, {Voutsinas}, {Weiler}, {Wevers}, {Wyrzykowski}, {Yoldas},
  {Yvard}, {Zhao}, {Zorec}, {Zucker}, {Zurbach}, \& {Zwitter}}]{gaia3}
{Gaia Collaboration}, {Brown}, A.~G.~A., {Vallenari}, A., {et~al.} 2021, \aap,
  649, A1

\bibitem[{{Gaia Collaboration} {et~al.}(2016){Gaia Collaboration}, {Prusti},
  {de Bruijne}, {Brown}, {Vallenari}, {Babusiaux}, {Bailer-Jones}, {Bastian},
  {Biermann}, {Evans}, \& et~al.}]{gaia1}
{Gaia Collaboration}, {Prusti}, T., {de Bruijne}, J.~H.~J., {et~al.} 2016,
  \aap, 595, A1

\bibitem[{{Gaia Collaboration} {et~al.}(2023){Gaia Collaboration}, {Vallenari},
  {Brown}, {Prusti}, {de Bruijne}, {Arenou}, {Babusiaux}, {Biermann},
  {Creevey}, {Ducourant}, {Evans}, {Eyer}, {Guerra}, {Hutton}, {Jordi},
  {Klioner}, {Lammers}, {Lindegren}, {Luri}, {Mignard}, {Panem}, {Pourbaix},
  {Randich}, {Sartoretti}, {Soubiran}, {Tanga}, {Walton}, {Bailer-Jones},
  {Bastian}, {Drimmel}, {Jansen}, {Katz}, {Lattanzi}, {van Leeuwen}, {Bakker},
  {Cacciari}, {Casta{\~n}eda}, {De Angeli}, {Fabricius}, {Fouesneau},
  {Fr{\'e}mat}, {Galluccio}, {Guerrier}, {Heiter}, {Masana}, {Messineo},
  {Mowlavi}, {Nicolas}, {Nienartowicz}, {Pailler}, {Panuzzo}, {Riclet}, {Roux},
  {Seabroke}, {Sordo}, {Th{\'e}venin}, {Gracia-Abril}, {Portell}, {Teyssier},
  {Altmann}, {Andrae}, {Audard}, {Bellas-Velidis}, {Benson}, {Berthier},
  {Blomme}, {Burgess}, {Busonero}, {Busso}, {C{\'a}novas}, {Carry}, {Cellino},
  {Cheek}, {Clementini}, {Damerdji}, {Davidson}, {de Teodoro}, {Nu{\~n}ez
  Campos}, {Delchambre}, {Dell'Oro}, {Esquej}, {Fern{\'a}ndez-Hern{\'a}ndez},
  {Fraile}, {Garabato}, {Garc{\'\i}a-Lario}, {Gosset}, {Haigron}, {Halbwachs},
  {Hambly}, {Harrison}, {Hern{\'a}ndez}, {Hestroffer}, {Hodgkin}, {Holl},
  {Jan{\ss}en}, {Jevardat de Fombelle}, {Jordan}, {Krone-Martins}, {Lanzafame},
  {L{\"o}ffler}, {Marchal}, {Marrese}, {Moitinho}, {Muinonen}, {Osborne},
  {Pancino}, {Pauwels}, {Recio-Blanco}, {Reyl{\'e}}, {Riello}, {Rimoldini},
  {Roegiers}, {Rybizki}, {Sarro}, {Siopis}, {Smith}, {Sozzetti}, {Utrilla},
  {van Leeuwen}, {Abbas}, {{\'A}brah{\'a}m}, {Abreu Aramburu}, {Aerts},
  {Aguado}, {Ajaj}, {Aldea-Montero}, {Altavilla}, {{\'A}lvarez}, {Alves},
  {Anders}, {Anderson}, {Anglada Varela}, {Antoja}, {Baines}, {Baker},
  {Balaguer-N{\'u}{\~n}ez}, {Balbinot}, {Balog}, {Barache}, {Barbato},
  {Barros}, {Barstow}, {Bartolom{\'e}}, {Bassilana}, {Bauchet}, {Becciani},
  {Bellazzini}, {Berihuete}, {Bernet}, {Bertone}, {Bianchi}, {Binnenfeld},
  {Blanco-Cuaresma}, {Blazere}, {Boch}, {Bombrun}, {Bossini}, {Bouquillon},
  {Bragaglia}, {Bramante}, {Breedt}, {Bressan}, {Brouillet}, {Brugaletta},
  {Bucciarelli}, {Burlacu}, {Butkevich}, {Buzzi}, {Caffau}, {Cancelliere},
  {Cantat-Gaudin}, {Carballo}, {Carlucci}, {Carnerero}, {Carrasco},
  {Casamiquela}, {Castellani}, {Castro-Ginard}, {Chaoul}, {Charlot}, {Chemin},
  {Chiaramida}, {Chiavassa}, {Chornay}, {Comoretto}, {Contursi}, {Cooper},
  {Cornez}, {Cowell}, {Crifo}, {Cropper}, {Crosta}, {Crowley}, {Dafonte},
  {Dapergolas}, {David}, {David}, {de Laverny}, {De Luise}, {De March}, {De
  Ridder}, {de Souza}, {de Torres}, {del Peloso}, {del Pozo}, {Delbo},
  {Delgado}, {Delisle}, {Demouchy}, {Dharmawardena}, {Di Matteo}, {Diakite},
  {Diener}, {Distefano}, {Dolding}, {Edvardsson}, {Enke}, {Fabre}, {Fabrizio},
  {Faigler}, {Fedorets}, {Fernique}, {Fienga}, {Figueras}, {Fournier},
  {Fouron}, {Fragkoudi}, {Gai}, {Garcia-Gutierrez}, {Garcia-Reinaldos},
  {Garc{\'\i}a-Torres}, {Garofalo}, {Gavel}, {Gavras}, {Gerlach}, {Geyer},
  {Giacobbe}, {Gilmore}, {Girona}, {Giuffrida}, {Gomel}, {Gomez},
  {Gonz{\'a}lez-N{\'u}{\~n}ez}, {Gonz{\'a}lez-Santamar{\'\i}a},
  {Gonz{\'a}lez-Vidal}, {Granvik}, {Guillout}, {Guiraud},
  {Guti{\'e}rrez-S{\'a}nchez}, {Guy}, {Hatzidimitriou}, {Hauser}, {Haywood},
  {Helmer}, {Helmi}, {Sarmiento}, {Hidalgo}, {Hilger}, {H{\l}adczuk}, {Hobbs},
  {Holland}, {Huckle}, {Jardine}, {Jasniewicz}, {Jean-Antoine Piccolo},
  {Jim{\'e}nez-Arranz}, {Jorissen}, {Juaristi Campillo}, {Julbe}, {Karbevska},
  {Kervella}, {Khanna}, {Kontizas}, {Kordopatis}, {Korn}, {K{\'o}sp{\'a}l},
  {Kostrzewa-Rutkowska}, {Kruszy{\'n}ska}, {Kun}, {Laizeau}, {Lambert},
  {Lanza}, {Lasne}, {Le Campion}, {Lebreton}, {Lebzelter}, {Leccia}, {Leclerc},
  {Lecoeur-Taibi}, {Liao}, {Licata}, {Lindstr{\o}m}, {Lister}, {Livanou},
  {Lobel}, {Lorca}, {Loup}, {Madrero Pardo}, {Magdaleno Romeo}, {Managau},
  {Mann}, {Manteiga}, {Marchant}, {Marconi}, {Marcos}, {Marcos Santos},
  {Mar{\'\i}n Pina}, {Marinoni}, {Marocco}, {Marshall}, {Martin Polo},
  {Mart{\'\i}n-Fleitas}, {Marton}, {Mary}, {Masip}, {Massari},
  {Mastrobuono-Battisti}, {Mazeh}, {McMillan}, {Messina}, {Michalik}, {Millar},
  {Mints}, {Molina}, {Molinaro}, {Moln{\'a}r}, {Monari}, {Mongui{\'o}},
  {Montegriffo}, {Montero}, {Mor}, {Mora}, {Morbidelli}, {Morel}, {Morris},
  {Muraveva}, {Murphy}, {Musella}, {Nagy}, {Noval}, {Oca{\~n}a}, {Ogden},
  {Ordenovic}, {Osinde}, {Pagani}, {Pagano}, {Palaversa}, {Palicio},
  {Pallas-Quintela}, {Panahi}, {Payne-Wardenaar}, {Pe{\~n}alosa Esteller},
  {Penttil{\"a}}, {Pichon}, {Piersimoni}, {Pineau}, {Plachy}, {Plum}, {Poggio},
  {Pr{\v{s}}a}, {Pulone}, {Racero}, {Ragaini}, {Rainer}, {Raiteri}, {Rambaux},
  {Ramos}, {Ramos-Lerate}, {Re Fiorentin}, {Regibo}, {Richards}, {Rios Diaz},
  {Ripepi}, {Riva}, {Rix}, {Rixon}, {Robichon}, {Robin}, {Robin}, {Roelens},
  {Rogues}, {Rohrbasser}, {Romero-G{\'o}mez}, {Rowell}, {Royer}, {Ruz Mieres},
  {Rybicki}, {Sadowski}, {S{\'a}ez N{\'u}{\~n}ez}, {Sagrist{\`a} Sell{\'e}s},
  {Sahlmann}, {Salguero}, {Samaras}, {Sanchez Gimenez}, {Sanna},
  {Santove{\~n}a}, {Sarasso}, {Schultheis}, {Sciacca}, {Segol}, {Segovia},
  {S{\'e}gransan}, {Semeux}, {Shahaf}, {Siddiqui}, {Siebert}, {Siltala},
  {Silvelo}, {Slezak}, {Slezak}, {Smart}, {Snaith}, {Solano}, {Solitro},
  {Souami}, {Souchay}, {Spagna}, {Spina}, {Spoto}, {Steele},
  {Steidelm{\"u}ller}, {Stephenson}, {S{\"u}veges}, {Surdej}, {Szabados},
  {Szegedi-Elek}, {Taris}, {Taylor}, {Teixeira}, {Tolomei}, {Tonello}, {Torra},
  {Torra}, {Torralba Elipe}, {Trabucchi}, {Tsounis}, {Turon}, {Ulla}, {Unger},
  {Vaillant}, {van Dillen}, {van Reeven}, {Vanel}, {Vecchiato}, {Viala},
  {Vicente}, {Voutsinas}, {Weiler}, {Wevers}, {Wyrzykowski}, {Yoldas}, {Yvard},
  {Zhao}, {Zorec}, {Zucker}, \& {Zwitter}}]{gaiadr3}
{Gaia Collaboration}, {Vallenari}, A., {Brown}, A.~G.~A., {et~al.} 2023, \aap,
  674, A1

\bibitem[{{Geier}(2020)}]{subkatii}
{Geier}, S. 2020, \aap, 635, A193

\bibitem[{{Geier} {et~al.}(2022){Geier}, {Dorsch}, {Pelisoli}, {Reindl},
  {Heber}, \& {Irrgang}}]{geiradprom}
{Geier}, S., {Dorsch}, M., {Pelisoli}, I., {et~al.} 2022, \aap, 661, A113

\bibitem[{{Gormaz-Matamala} {et~al.}(2019){Gormaz-Matamala}, {Cur{\'e}},
  {Cidale}, \& {Venero}}]{gmmcak}
{Gormaz-Matamala}, A.~C., {Cur{\'e}}, M., {Cidale}, L.~S., \& {Venero},
  R.~O.~J. 2019, \apj, 873, 131

\bibitem[{{Gr{\"a}fener} \& {Hamann}(2008)}]{grahamz}
{Gr{\"a}fener}, G. \& {Hamann}, W.-R. 2008, \aap, 482, 945

\bibitem[{{Guo}(2018)}]{guo}
{Guo}, J.-J. 2018, \apj, 866, 58

\bibitem[{{Hall} \& {Jeffery}(2016)}]{hallvodik}
{Hall}, P.~D. \& {Jeffery}, C.~S. 2016, \mnras, 463, 2756

\bibitem[{{Han} {et~al.}(2003){Han}, {Podsiadlowski}, {Maxted}, \&
  {Marsh}}]{hanhehe}
{Han}, Z., {Podsiadlowski}, P., {Maxted}, P.~F.~L., \& {Marsh}, T.~R. 2003,
  \mnras, 341, 669

\bibitem[{{Han} {et~al.}(2002){Han}, {Podsiadlowski}, {Maxted}, {Marsh}, \&
  {Ivanova}}]{han}
{Han}, Z., {Podsiadlowski}, P., {Maxted}, P.~F.~L., {Marsh}, T.~R., \&
  {Ivanova}, N. 2002, \mnras, 336, 449

\bibitem[{{Heap} {et~al.}(2006){Heap}, {Lanz}, \& {Hubeny}}]{kupa}
{Heap}, S.~R., {Lanz}, T., \& {Hubeny}, I. 2006, \apj, 638, 409

\bibitem[{{Heber}(2016)}]{heberpreh}
{Heber}, U. 2016, \pasp, 128, 082001

\bibitem[{{Hirsch}(2009)}]{lepsiindex}
{Hirsch}, H.~A. 2009, PhD thesis, Friedrich Alexander University of
  Erlangen-Nuremberg, Germany

\bibitem[{{H{\o}g} {et~al.}(2000){H{\o}g}, {Fabricius}, {Makarov}, {Urban},
  {Corbin}, {Wycoff}, {Bastian}, {Schwekendiek}, \& {Wicenec}}]{hogefabe}
{H{\o}g}, E., {Fabricius}, C., {Makarov}, V.~V., {et~al.} 2000, \aap, 355, L27

\bibitem[{{Hubeny}(1988)}]{tlusty0}
{Hubeny}, I. 1988, Computer Physics Communications, 52, 103

\bibitem[{{Hubeny} \& {Lanz}(2011)}]{synspec}
{Hubeny}, I. \& {Lanz}, T. 2011, {Synspec: General Spectrum Synthesis Program},
  Astrophysics Source Code Library

\bibitem[{{Hummer} {et~al.}(1993){Hummer}, {Berrington}, {Eissner}, {Pradhan},
  {Saraph}, \& {Tully}}]{zel0}
{Hummer}, D.~G., {Berrington}, K.~A., {Eissner}, W., {et~al.} 1993, \aap, 279,
  298

\bibitem[{{Husfeld} {et~al.}(1989){Husfeld}, {Butler}, {Heber}, \&
  {Drilling}}]{sam15}
{Husfeld}, D., {Butler}, K., {Heber}, U., \& {Drilling}, J.~S. 1989, \aap, 222,
  150

\bibitem[{{Jeffery} {et~al.}(2011){Jeffery}, {Karakas}, \& {Saio}}]{jefkas}
{Jeffery}, C.~S., {Karakas}, A.~I., \& {Saio}, H. 2011, \mnras, 414, 3599

\bibitem[{{Justham} {et~al.}(2011){Justham}, {Podsiadlowski}, \&
  {Han}}]{jencohe}
{Justham}, S., {Podsiadlowski}, P., \& {Han}, Z. 2011, \mnras, 410, 984

\bibitem[{{Klement} {et~al.}(2022){Klement}, {Schaefer}, {Gies}, {Wang},
  {Baade}, {Rivinius}, {Gallenne}, {Carciofi}, {Monnier}, {M{\'e}rand},
  {Anugu}, {Kraus}, {Davies}, {Lanthermann}, {Gardner}, {Wysocki}, {Ennis},
  {Labdon}, {Setterholm}, \& {Le Bouquin}}]{robpod}
{Klement}, R., {Schaefer}, G.~H., {Gies}, D.~R., {et~al.} 2022, \apj, 926, 213

\bibitem[{{Koubsk{\'y}} {et~al.}(2014){Koubsk{\'y}}, {Kotkov{\'a}}, {Kraus},
  {Yang}, {{\v{S}}lechta}, {Harmanec}, {Wolf}, {Votruba}, {Kub{\'a}t},
  {Kub{\'a}tov{\'a}}, {Niemczura}, \& {{\v{S}}koda}}]{kourad}
{Koubsk{\'y}}, P., {Kotkov{\'a}}, L., {Kraus}, M., {et~al.} 2014, \aap, 567,
  A57

\bibitem[{{Kramer} {et~al.}(2020){Kramer}, {Schneider}, {Ohlmann}, {Geier},
  {Schaffenroth}, {Pakmor}, \& {R{\"o}pke}}]{kramuvol}
{Kramer}, M., {Schneider}, F.~R.~N., {Ohlmann}, S.~T., {et~al.} 2020, \aap,
  642, A97

\bibitem[{{Kramida} {et~al.}(2015){Kramida}, {Ralchenko}, {Reader}, \& {NIST
  ASD Team}}]{nist}
{Kramida}, A., {Ralchenko}, Y., {Reader}, J., \& {NIST ASD Team}. 2015, {NIST
  Atomic Spectra Database},
  \url{https://www.nist.gov/pml/atomic-spectra-database}, [Online]

\bibitem[{{Krti{\v c}ka} \& {Kub{\'a}t}(2006)}]{bezvi}
{Krti{\v c}ka}, J. \& {Kub{\'a}t}, J. 2006, \aap, 446, 1039

\bibitem[{{Krti{\v c}ka} \& {Kub{\'a}t}(2009)}]{cnovit}
{Krti{\v c}ka}, J. \& {Kub{\'a}t}, J. 2009, \aap, 493, 585

\bibitem[{{Krti{\v c}ka} \& {Kub{\'a}t}(2017)}]{cmfkont}
{Krti{\v c}ka}, J. \& {Kub{\'a}t}, J. 2017, \aap, 606, A31

\bibitem[{{Krti{\v c}ka} {et~al.}(2016){Krti{\v c}ka}, {Kub{\'a}t}, \& {Krti{\v
  c}kov{\'a}}}]{snehurka}
{Krti{\v c}ka}, J., {Kub{\'a}t}, J., \& {Krti{\v c}kov{\'a}}, I. 2016, \aap,
  593, A101

\bibitem[{{Krti{\v c}ka} \& {{\v S}tefl}(1999)}]{kobr}
{Krti{\v c}ka}, J. \& {{\v S}tefl}, V. 1999, \aaps, 138, 47

\bibitem[{{Krti{\v{c}}ka} {et~al.}(2019){Krti{\v{c}}ka}, {Jan{\'\i}k},
  {Krti{\v{c}}kov{\'a}}, {Mereghetti}, {Pintore}, {N{\'e}meth}, {Kub{\'a}t}, \&
  {Vu{\v{c}}kovi{\'c}}}]{esosubwind}
{Krti{\v{c}}ka}, J., {Jan{\'\i}k}, J., {Krti{\v{c}}kov{\'a}}, I., {et~al.}
  2019, \aap, 631, A75

\bibitem[{{Krti{\v{c}}ka} {et~al.}(2020){Krti{\v{c}}ka}, {Kub{\'a}t}, \&
  {Krti{\v{c}}kov{\'a}}}]{btvit}
{Krti{\v{c}}ka}, J., {Kub{\'a}t}, J., \& {Krti{\v{c}}kov{\'a}}, I. 2020, \aap,
  635, A173

\bibitem[{{Kupka} {et~al.}(1999){Kupka}, {Piskunov}, {Ryabchikova}, {Stempels},
  \& {Weiss}}]{vald2}
{Kupka}, F., {Piskunov}, N., {Ryabchikova}, T.~A., {Stempels}, H.~C., \&
  {Weiss}, W.~W. 1999, \aaps, 138, 119

\bibitem[{{La Palombara} {et~al.}(2014){La Palombara}, {Esposito},
  {Mereghetti}, \& {Tiengo}}]{bufacek}
{La Palombara}, N., {Esposito}, P., {Mereghetti}, S., \& {Tiengo}, A. 2014,
  \aap, 566, A4

\bibitem[{{Lanz} {et~al.}(2004){Lanz}, {Brown}, {Sweigart}, {Hubeny}, \&
  {Landsman}}]{sam4}
{Lanz}, T., {Brown}, T.~M., {Sweigart}, A.~V., {Hubeny}, I., \& {Landsman},
  W.~B. 2004, \apj, 602, 342

\bibitem[{{Lanz} \& {Hubeny}(2003)}]{ostar2003}
{Lanz}, T. \& {Hubeny}, I. 2003, \apjs, 146, 417

\bibitem[{{Latour} {et~al.}(2018){Latour}, {Chayer}, {Green}, {Irrgang}, \&
  {Fontaine}}]{lathbt}
{Latour}, M., {Chayer}, P., {Green}, E.~M., {Irrgang}, A., \& {Fontaine}, G.
  2018, \aap, 609, A89

\bibitem[{{Lucy} \& {Solomon}(1970)}]{lusol}
{Lucy}, L.~B. \& {Solomon}, P.~M. 1970, \apj, 159, 879

\bibitem[{{Luo} {et~al.}(2019){Luo}, {N{\'e}meth}, {Deng}, \&
  {Han}}]{luogailam}
{Luo}, Y., {N{\'e}meth}, P., {Deng}, L., \& {Han}, Z. 2019, \apj, 881, 7

\bibitem[{{Maeder}(2009)}]{biblerot}
{Maeder}, A. 2009, {Physics, Formation and Evolution of Rotating Stars}

\bibitem[{{Marcolino} {et~al.}(2022){Marcolino}, {Bouret}, {Rocha-Pinto},
  {Bernini-Peron}, \& {Vink}}]{marcoz}
{Marcolino}, W.~L.~F., {Bouret}, J.~C., {Rocha-Pinto}, H.~J., {Bernini-Peron},
  M., \& {Vink}, J.~S. 2022, \mnras, 511, 5104

\bibitem[{{Mereghetti} \& {La Palombara}(2016)}]{sandroprehled}
{Mereghetti}, S. \& {La Palombara}, N. 2016, Advances in Space Research, 58,
  809

\bibitem[{{Michaud} {et~al.}(2011){Michaud}, {Richer}, \& {Richard}}]{miriri}
{Michaud}, G., {Richer}, J., \& {Richard}, O. 2011, \aap, 529, A60

\bibitem[{{Miller Bertolami} {et~al.}(2022){Miller Bertolami}, {Battich},
  {C{\'o}rsico}, {Althaus}, \& {Wachlin}}]{milbicohe}
{Miller Bertolami}, M.~M., {Battich}, T., {C{\'o}rsico}, A.~H., {Althaus},
  L.~G., \& {Wachlin}, F.~C. 2022, \mnras, 511, L60

\bibitem[{Naslim {et~al.}(2019)Naslim, Jeffery, \& Woolf}]{nastez}
Naslim, N., Jeffery, C.~S., \& Woolf, V.~M. 2019, Monthly Notices of the Royal
  Astronomical Society, 491, 874

\bibitem[{{Naz{\'e}}(2009)}]{naze}
{Naz{\'e}}, Y. 2009, \aap, 506, 1055

\bibitem[{{N{\'e}meth} {et~al.}(2012){N{\'e}meth}, {Kawka}, \&
  {Vennes}}]{nemgalex}
{N{\'e}meth}, P., {Kawka}, A., \& {Vennes}, S. 2012, \mnras, 427, 2180

\bibitem[{{Owocki} {et~al.}(2013){Owocki}, {Sundqvist}, {Cohen}, \&
  {Gayley}}]{owomix}
{Owocki}, S.~P., {Sundqvist}, J.~O., {Cohen}, D.~H., \& {Gayley}, K.~G. 2013,
  \mnras, 429, 3379

\bibitem[{{Palacios} {et~al.}(2010){Palacios}, {Gebran}, {Josselin}, {Martins},
  {Plez}, {Belmas}, \& {L{\`e}bre}}]{pollux}
{Palacios}, A., {Gebran}, M., {Josselin}, E., {et~al.} 2010, \aap, 516, A13

\bibitem[{{Paunzen}(2015)}]{ernstuvby}
{Paunzen}, E. 2015, \aap, 580, A23

\bibitem[{{Pelisoli} {et~al.}(2020){Pelisoli}, {Vos}, {Geier}, {Schaffenroth},
  \& {Baran}}]{saminesami}
{Pelisoli}, I., {Vos}, J., {Geier}, S., {Schaffenroth}, V., \& {Baran}, A.~S.
  2020, \aap, 642, A180

\bibitem[{{Peters} {et~al.}(2008){Peters}, {Gies}, {Grundstrom}, \&
  {McSwain}}]{petfycan}
{Peters}, G.~J., {Gies}, D.~R., {Grundstrom}, E.~D., \& {McSwain}, M.~V. 2008,
  \apj, 686, 1280

\bibitem[{{Piskunov} {et~al.}(1995){Piskunov}, {Kupka}, {Ryabchikova}, {Weiss},
  \& {Jeffery}}]{vald1}
{Piskunov}, N.~E., {Kupka}, F., {Ryabchikova}, T.~A., {Weiss}, W.~W., \&
  {Jeffery}, C.~S. 1995, \aaps, 112, 525

\bibitem[{{Pols} {et~al.}(1991){Pols}, {Cote}, {Waters}, \& {Heise}}]{bepols}
{Pols}, O.~R., {Cote}, J., {Waters}, L.~B.~F.~M., \& {Heise}, J. 1991, \aap,
  241, 419

\bibitem[{{Puls}(2008)}]{pulboh}
{Puls}, J. 2008, in The Metal-Rich Universe, ed. G.~{Israelian} \& G.~{Meynet},
  295

\bibitem[{{Saio} \& {Jeffery}(2000)}]{saje}
{Saio}, H. \& {Jeffery}, C.~S. 2000, \mnras, 313, 671

\bibitem[{{Sander} {et~al.}(2017){Sander}, {Hamann}, {Todt}, {Hainich}, \&
  {Shenar}}]{powrdyn}
{Sander}, A.~A.~C., {Hamann}, W.-R., {Todt}, H., {Hainich}, R., \& {Shenar}, T.
  2017, \aap, 603, A86

\bibitem[{{Sander} {et~al.}(2023){Sander}, {Lefever}, {Poniatowski},
  {Ramachandran}, {Sabhahit}, \& {Vink}}]{wrtep}
{Sander}, A.~A.~C., {Lefever}, R.~R., {Poniatowski}, L.~G., {et~al.} 2023,
  \aap, 670, A83

\bibitem[{{Schindewolf} {et~al.}(2018){Schindewolf}, {N{\'e}meth}, {Heber},
  {Battich}, {Miller Bertolami}, {Irrgang}, \& {Latour}}]{svlk}
{Schindewolf}, M., {N{\'e}meth}, P., {Heber}, U., {et~al.} 2018, \aap, 620, A36

\bibitem[{{Sch\"onberner} \& {Drilling}(1984)}]{peknevrtani}
{Sch\"onberner}, D. \& {Drilling}, J.~S. 1984, \apj, 278, 702

\bibitem[{{Seaton} {et~al.}(1992){Seaton}, {Zeippen}, {Tully}, {Pradhan},
  {Mendoza}, {Hibbert}, \& {Berrington}}]{topt}
{Seaton}, M.~J., {Zeippen}, C.~J., {Tully}, J.~A., {et~al.} 1992, \rmxaa, 23

\bibitem[{Shao \& Li(2021)}]{shaolin}
Shao, Y. \& Li, X.-D. 2021, The Astrophysical Journal, 908, 67

\bibitem[{{Skrutskie} {et~al.}(2006){Skrutskie}, {Cutri}, {Stiening},
  {Weinberg}, {Schneider}, {Carpenter}, {Beichman}, {Capps}, {Chester},
  {Elias}, {Huchra}, {Liebert}, {Lonsdale}, {Monet}, {Price}, {Seitzer},
  {Jarrett}, {Kirkpatrick}, {Gizis}, {Howard}, {Evans}, {Fowler}, {Fullmer},
  {Hurt}, {Light}, {Kopan}, {Marsh}, {McCallon}, {Tam}, {Van Dyk}, \&
  {Wheelock}}]{2mass}
{Skrutskie}, M.~F., {Cutri}, R.~M., {Stiening}, R., {et~al.} 2006, \aj, 131,
  1163

\bibitem[{{Sundqvist} {et~al.}(2019){Sundqvist}, {Bj{\"o}rklund}, {Puls}, \&
  {Najarro}}]{sundyn}
{Sundqvist}, J.~O., {Bj{\"o}rklund}, R., {Puls}, J., \& {Najarro}, F. 2019,
  \aap, 632, A126

\bibitem[{{Sundqvist} {et~al.}(2010){Sundqvist}, {Puls}, \&
  {Feldmeier}}]{chuchcar}
{Sundqvist}, J.~O., {Puls}, J., \& {Feldmeier}, A. 2010, \aap, 510, A11

\bibitem[{{Unglaub} \& {Bues}(2001)}]{unbuhb}
{Unglaub}, K. \& {Bues}, I. 2001, \aap, 374, 570

\bibitem[{{{\v S}urlan} {et~al.}(2012){{\v S}urlan}, {Hamann}, {Kub{\'a}t},
  {Oskinova}, \& {Feldmeier}}]{clres1}
{{\v S}urlan}, B., {Hamann}, W.-R., {Kub{\'a}t}, J., {Oskinova}, L.~M., \&
  {Feldmeier}, A. 2012, \aap, 541, A37

\bibitem[{{Vauclair}(1975)}]{vasam}
{Vauclair}, S. 1975, \aap, 45, 233

\bibitem[{{Vauclair} {et~al.}(1991){Vauclair}, {Dolez}, \& {Gough}}]{vadog}
{Vauclair}, S., {Dolez}, N., \& {Gough}, D.~O. 1991, \aap, 252, 618

\bibitem[{{Vink} \& {Cassisi}(2002)}]{vinca}
{Vink}, J.~S. \& {Cassisi}, S. 2002, \aap, 392, 553

\bibitem[{{Vink} \& {de Koter}(2002)}]{vinkolbv}
{Vink}, J.~S. \& {de Koter}, A. 2002, \aap, 393, 543

\bibitem[{{Werner} {et~al.}(2022){Werner}, {Reindl}, {Geier}, \&
  {Pritzkuleit}}]{werash}
{Werner}, K., {Reindl}, N., {Geier}, S., \& {Pritzkuleit}, M. 2022, \mnras,
  511, L66

\bibitem[{{Wright} {et~al.}(2010){Wright}, {Eisenhardt}, {Mainzer}, {Ressler},
  {Cutri}, {Jarrett}, {Kirkpatrick}, {Padgett}, {McMillan}, {Skrutskie},
  {Stanford}, {Cohen}, {Walker}, {Mather}, {Leisawitz}, {Gautier}, {McLean},
  {Benford}, {Lonsdale}, {Blain}, {Mendez}, {Irace}, {Duval}, {Liu}, {Royer},
  {Heinrichsen}, {Howard}, {Shannon}, {Kendall}, {Walsh}, {Larsen}, {Cardon},
  {Schick}, {Schwalm}, {Abid}, {Fabinsky}, {Naes}, \& {Tsai}}]{vosawise}
{Wright}, E.~L., {Eisenhardt}, P. R.~M., {Mainzer}, A.~K., {et~al.} 2010, \aj,
  140, 1868

\bibitem[{{Xiong} {et~al.}(2017){Xiong}, {Chen}, {Podsiadlowski}, {Li}, \&
  {Han}}]{xiong}
{Xiong}, H., {Chen}, X., {Podsiadlowski}, P., {Li}, Y., \& {Han}, Z. 2017,
  \aap, 599, A54

\bibitem[{{Zhang} {et~al.}(2017){Zhang}, {Hall}, {Jeffery}, \& {Bi}}]{splybthp}
{Zhang}, X., {Hall}, P.~D., {Jeffery}, C.~S., \& {Bi}, S. 2017, \apj, 835, 242

\bibitem[{{Zhang} \& {Jeffery}(2012{\natexlab{a}})}]{zhaj}
{Zhang}, X. \& {Jeffery}, C.~S. 2012{\natexlab{a}}, \mnras, 426, L81

\bibitem[{{Zhang} \& {Jeffery}(2012{\natexlab{b}})}]{zhaff}
{Zhang}, X. \& {Jeffery}, C.~S. 2012{\natexlab{b}}, \mnras, 419, 452

\bibitem[{{Zverko} {et~al.}(2007){Zverko}, {{\v Z}i{\v z}{\v n}ovsk{\'y}},
  {Mikul{\'a}{\v s}ek}, \& {Iliev}}]{zvezimi}
{Zverko}, J., {{\v Z}i{\v z}{\v n}ovsk{\'y}}, J., {Mikul{\'a}{\v s}ek}, Z., \&
  {Iliev}, I.~K. 2007, Contributions of the Astronomical Observatory Skalnate
  Pleso, 37, 49

\end{thebibliography}
\end{document}